\documentclass[11pt]{article}

\usepackage{amssymb,amsmath,amsthm,amsfonts,amstext}

\usepackage[title]{appendix}

\usepackage[left=1.5cm,right=1.5cm,top=1.5cm,bottom=1.5cm]{geometry}

\usepackage{graphicx}
\usepackage[colorlinks=true, allcolors=blue]{hyperref}

\usepackage[english]{babel}

\usepackage{enumerate}

\usepackage[mathscr]{euscript}

\usepackage{thm-restate}

\usepackage{orcidlink}

\newcommand\IGNORE[1]{}

\newtheorem{theorem}{Theorem}
\newtheorem{corollary}[theorem]{Corollary}
\newtheorem{proposition}[theorem]{Proposition}
\newtheorem{lemma}[theorem]{Lemma}

\newtheorem{claim}[theorem]{Claim}
\newtheorem*{remark}{Remark}

\newtheorem{example}{Example}

\usepackage[ruled]{algorithm2e}
\usepackage{tikz}
\usetikzlibrary{decorations}
\usetikzlibrary{decorations.pathreplacing}

\newcommand\lpopt{\hbox{\text{LP}$_{\text{opt}}$}}

\newcommand{\Q}{\ensuremath{\mathbb Q}}
\newcommand{\Qp}{\ensuremath{\Q_{\geq 0}}}
\newcommand{\Zint}{\ensuremath{\mathbb Z}}
\newcommand{\Zp}{\ensuremath{\Zint_{\geq 0}}}

\newcommand{\opt}{\textsc{opt}}
\newcommand{\safe}{\mathscr{S}}
\newcommand{\unsafe}{\mathscr{U}}

\newcommand{\fgc}{\mathrm{FGC}}
\newcommand{\pqfgc}{(p,q)\text{-}\fgc}
\newcommand{\ponefgc}{(p,1)\text{-}\fgc}
\newcommand{\ptwofgc}{(p,2)\text{-}\fgc}
\newcommand{\pthreefgc}{(p,3)\text{-}\fgc}

\newcommand{\oneqfgc}{(1,q)\text{-}\fgc}
\newcommand{\oneqplusfgc}{(1,q+1)\text{-}\fgc}

\newcommand{\C}{\mathscr{C}}	%% script C
\newcommand{\F}{\mathcal{F}}	%% script F
\newcommand{\J}{{J}}		%% edge-set J

\newcommand{\alg}{\textsc{alg}}

\newcommand{\twoqfgc}{(2,q)\text{-}\fgc}

\newcommand\capbound{\lambda}

\newcommand\hG{\hat{G}}
\newcommand\hE{\hat{E}}
\newcommand\hV{\hat{V}}

\newcommand\tG{\widetilde{G}}
\newcommand\tE{\widetilde{E}}
\newcommand\tV{\widetilde{V}}

\newcommand{\pliableapx}{5}
\newcommand{\pliableapxOLD}{6}

\newcommand{\oneqfgcapx}{7}

\newcommand\ASC{\mathrm{Cover\,Small\,Cuts}}
\newcommand\twoASC{\mathrm{2\text{-}Cover\,Small\,Cuts}}
\newcommand\threeASC{\mathrm{3\text{-}Cover\,Small\,Cuts}}

\newcommand{\ds}{\displaystyle}

\newcommand{\ifinal}{\ensuremath{i_{\textsl{final}}}}

\newcommand{\Lcur}{\ensuremath{{\ell}_{\textsl{c}}}}
\newcommand{\Hcur}{\ensuremath{{h}_{\textsl{c}}}}
\newcommand{\Lfinal}{\ensuremath{{\ell}_{\textsl{final}}}}
\newcommand{\Hfinal}{\ensuremath{{h}_{\textsl{final}}}}

\title{Improved Approximation Algorithms for Capacitated Network Design and Flexible Graph Connectivity}

\author{\large
Ishan Bansal\thanks{
    Email: \href{mailto:ib332@cornell.edu}{\texttt{ib332@cornell.edu}}. 
    ORCID: \orcidlink{0000-0002-5083-309X} \href{https://orcid.org/0000-0002-5083-309X}{0000-0002-5083-309X}. 
	Amazon, Bellevue, WA, USA. This work is external and does not relate to the position at Amazon. }
\and
Joseph Cheriyan\thanks{
    Email: \href{mailto:jcheriyan@uwaterloo.ca}{\texttt{jcheriyan@uwaterloo.ca}}. 
    Web: \url{https://www.math.uwaterloo.ca/~jcheriyan}. 
    ORCID: \orcidlink{0000-0003-0316-7650} \href{https://orcid.org/0000-0003-0316-7650}{0000-0003-0316-7650}. 
    Department of Combinatorics \& Optimization, University of Waterloo, Canada. 
    Supported in part by NSERC grant RGPIN-2024-04473.}
\and
Sanjeev Khanna\thanks{
    Email: \href{mailto:sanjeev@cis.upenn.edu}{\texttt{sanjeev@cis.upenn.edu}}. 
    ORCID: \orcidlink{0009-0000-2601-1689} \href{https://orcid.org/0009-0000-2601-1689}{0009-0000-2601-1689}. 
	{Department of Computer and Information Science, University of Pennsylvania,
		Philadelphia, USA.}
	Supported in part by NSF award CCF-2402284 and AFOSR award FA9550-25-1-0107.}
\and 
Miles Simmons\thanks{
    Email: \href{mailto:mjsimmons@uwaterloo.ca}{\texttt{mjsimmons@uwaterloo.ca}}. 
    ORCID: \orcidlink{0009-0006-6144-9735} \href{https://orcid.org/0009-0006-6144-9735}{0009-0006-6144-9735}. 
        Department of Combinatorics \& Optimization, University of Waterloo, Canada.}
}

\begin{document}

\maketitle

{
\begin{abstract}
We present improved approximation algorithms for some problems in the related areas
of Capacitated Network Design and Flexible Graph Connectivity.

In the Cap-$k$-ECSS problem, we are given a graph $G=(V,E)$ whose
edges have non-negative costs and positive integer capacities, and
the goal is to find a minimum-cost edge-set $F$ such that every
non-trivial cut of the graph $G'=(V,F)$ has capacity at least $k$.
We present an $O(\log{k})$-approximation algorithm for the Cap-$k$-ECSS
problem, asymptotically improving upon the previous best approximation
ratio of $\min(O(\log{n}),\; O(k))$ whenever $\log({k})=o(\log{n})$,
where $n$ denotes $|V|$. (Here, we are giving an overview of the previous
approximation ratios; section~\ref{sec:intro} has a detailed discussion.)

In the $(p,q)$-Flexible Graph Connectivity problem, denoted $\pqfgc$,
the input is a graph $G=(V, E)$ where $E$ is partitioned into
\textit{safe} and \textit{unsafe} edges, and the goal is to find a
minimum-cost edge-set $F$ such that the subgraph $G'=(V, F)$ remains
$p$-edge connected upon removal of any $q$ unsafe edges from $F$.
We present a $\oneqfgcapx$-approximation algorithm for the $\oneqfgc$
problem that improves upon the previous best approximation ratio of $(q+1)$.

Both of our results are obtained by using natural LP relaxations
strengthened with the knapsack-cover inequalities, and then, during
the rounding process, utilizing a recent $O(1)$-approximation
algorithm for the $\ASC$ problem. In the latter problem, the goal
is to find a minimum-cost set of links such that each non-trivial
cut of capacity less than a specified value is covered by a link.
We also show that the problem of covering small cuts inherently
arises in another variant of $\pqfgc$. Specifically, we give Cook
reductions that preserve approximation ratios within $O(1)$ factors
between the $\twoqfgc$ problem and the $\twoASC$ problem; in the
latter problem, each small cut needs to be covered by two links.
\end{abstract}
}

\vspace{1em} % Adds some vertical space

\noindent \textbf{ACM CCS Concepts:}
{Theory of computation $\longrightarrow$
	Approximation algorithms analysis,
	Network optimization}
\vspace{1em} % Adds some vertical space

\newpage

%%%%%%%%%%
{
\section{Introduction \label{sec:intro}}

Given a graph $G=(V,E)$ whose edges have both costs and capacities,
a fundamental task in network design is to find a spanning subgraph
of minimum cost that satisfies some specified connectivity requirements.
In this paper, we present results on two well-studied problems in the
area of approximation algorithms pertaining to the design of networks
with edge-connectivity requirements.

When all edges have the same capacity, a seminal result by
Jain~\cite{Jain01} provides a $2$-approximation algorithm for the
survivable network design problem (SNDP);
see section~\ref{sec:f-conn} for further discussion.

In the more general setting, where edges have non-uniform capacities,
the problem, referred to as {\em capacitated network design}, becomes more challenging.
Our focus is on designing approximation algorithms for the special
case called the Cap-$k$-ECSS problem where we are given a graph
$G=(V,E)$ with a non-negative cost and a positive integer capacity
for each edge, and the algorithmic goal is to find a minimum-cost
spanning subgraph such that every non-trivial cut has capacity $k$ or more.
Let $n=|V|$ and let $u_{min}$ (respectively, $u_{max}$) denote the
minimum (respectively, maximum) capacity of an edge; assume that $u_{max} \leq k$.
One of the earliest approximation algorithms for the Cap-$k$-ECSS
problem is due to Goemans et~al.~\cite{GGPSTW94}, and it achieves
an approximation ratio of $\min\{2k,|E|\}$.  The best approximation
ratio known is
$\min(O(\log{n}), k, 2u_{max}, \pliableapxOLD \cdot {\lceil k/u_{min} \rceil})$,
due to
Goemans et~al.~\cite{GGPSTW94},
Chakrabarty et~al.~\cite{CCKK15},
Boyd et~al.~\cite{BCHI24},
and Bansal \cite{B2023,B:ipco2025}.
Moreover, there are no known hardness results that rule out the
possibility of better asymptotic approximation ratios.

Another line of research called {\em flexible graph connectivity}
(abbreviated as FGC), has emerged recently, motivated by natural
questions in network design in the setting of robust optimization.
Adjiashvili, Hommelsheim and M\"uhlenthaler \cite{AHM22} proposed
an FGC model that distinguishes between safe (never-failing) and
unsafe (failure-prone) edges. The algorithmic goal is to choose a
set of edges of minimum~cost that satisfies a (global) edge-connectivity
requirement, while tolerating failures of up to a specified number
of unsafe edges. A basic problem in this setting is the $\oneqfgc$
problem where the goal is to ensure that the network remains connected
even after the failure of up to $q$ unsafe edges. We mention that
the $\oneqfgc$ problem can be modeled as a special case of the
Cap-$k$-ECSS problem.

In the rest of the introduction section, we first discuss related work, and
then we formalize the problems studied in this paper.
After that, we state our main results.

We may use abbreviations for some standard terms, e.g., we may use
``$\oneqfgc$'' as an abbreviation for ``the $\oneqfgc$ problem''.
For each of the minimization problems (in network design or flexible graph connectivity),
we use $\opt$ to denote the optimal value
(i.e., the minimum cost of an integer solution), and
$\lpopt$ to denote the optimal value of an LP relaxation.
The context will resolve potential ambiguities.

\subsection{Related Work}

As mentioned above, research on approximation algorithms for
Cap-$k$-ECSS was initiated by Goemans et~al.~\cite{GGPSTW94}.  Carr
et~al.~\cite{CFLP00}, in a seminal paper, introduced a key algorithmic
tool for capacitated network design based on the 
Knapsack-Cover Inequalities (KCI); we discuss KCI in more detail
in section~\ref{sec:KCI}.  More than a decade later, Chakrabarty
et~al.~\cite{CCKK15} used KCI to design an $O(\log{n})$ approximation
algorithm for Cap-$k$-ECSS. Boyd et~al.~\cite{BCHI24} gave a
$\min\{k,\;2u_{max}\}$-approximation algorithm for Cap-$k$-ECSS, and
Bansal \cite{B2023,B:ipco2025}, based on previous work by Bansal et~al.~\cite{BCGI24} and Williamson et~al.~\cite{WGMV95},
gave a $(\pliableapxOLD\cdot {\lceil k/u_{min} \rceil})$-approximation
algorithm for Cap-$k$-ECSS.

The model of flexible graph connectivity originated from research
in the area of robust optimization. Adjiashvili, Stiller and Zenklusen
\cite{ASZ15} introduced their model of bulk-robust combinatorial
optimization, and designed some approximation algorithms. Later,
Adjiashvili, Hommelsheim and M\"uhlenthaler \cite{AHM22} introduced
the $\fgc$ model. Boyd et~al.~\cite{BCHI24} introduced a generalization
called the $\pqfgc$ model.
Boyd et~al.~\cite{BCHI24} presented a $4$-approximation algorithm
for $\ponefgc$ based on the
primal-dual method of Williamson, Goemans, Mihail \& Vazirani (WGMV) \cite{WGMV95}, and a
$(q+1)$-approximation algorithm for $\oneqfgc$; moreover, they gave
an $O(q \log n)$-approximation algorithm for $\pqfgc$.
Subsequently, several interesting results and approximation algorithms have been presented;
we summarize some of these recent papers in chronological order.
Chekuri and Jain \cite{CJ23} give $O(p)$-approximation algorithms
for, respectively, $(p, 2)$, $(p, 3)$ and $(2p, 4)$-FGC,
and an $O(q)$-approximation algorithm for $\twoqfgc$. Bansal
et~al.~\cite{BCGI24}, among other results, give an $O(1)$-approximation
algorithm for $\ptwofgc$; moreover, they give a 16-approximation algorithm
for a related problem called $\ASC$. Nutov \cite{N:waoa2024} improves the
approximation ratio for $\ASC$ from 16 to 10.
Bansal \cite{B2023,B:ipco2025}, and later Nutov \cite{N:mfcs2025}, give a
$\pliableapxOLD$-approximation algorithm for $\ASC$; also, see \cite{SBC2025}.
Recently, Simmons et~al.~\cite{S2025,SBC-5approx} improved the
approximation ratio for $\ASC$ to $\pliableapx$.
Bansal \cite{B2023,B:ipco2025} gives an $O(1)$-approximation algorithm for $\pthreefgc$.
Hyatt-Denesik et~al.~\cite{HJS:esa2024}, among other results, give
approximation algorithms for unit-cost FGC problems with edge-connectivity
requirements as well as for unit-cost FGC problems with vertex
connectivity requirements.  Hommelsheim et~al.~\cite{HLMZ:stacs2025} study
a model related to a generalization of FGC called the
$(p,q)$-Steiner-Connectivity Preservation problem.  Ibrahimpur \&
Vegh \cite{IV:ipco2025} give an $O(\log{n})$-approximation algorithm for $\pqfgc$.

\subsection{Capacitated Network Design and the \texorpdfstring{Cap-$k$-ECSS}{Cap-k-ECSS} problem \label{sec:intro:cnd}}
{
The Cap-$k$-ECSS problem is as follows:
Given an undirected graph $G = (V,E)$ with edge costs $c \in \Qp^E$ and
edge capacities $u \in \Zint_{\geq 1}^E$, find a minimum-cost
edge-set $F\subseteq E$ such that the capacity of any
cut in $(V,F)$ is at least $k$.
Let $u_{min}$ (respectively, $u_{max}$) denote the minimum (respectively,
maximum) capacity of an edge in $E$, and assume (w.l.o.g.) that $u_{max} \leq k$.

For a graph $G=(V,E)$ and a set of nodes $S\subseteq{V}$, the
\textit{cut of $S$}, denoted by $\delta(S)$, refers to the set of
edges that have exactly one end-node in $S$.
Whenever we use the term ``cut $\delta(S)$'' we mean that $S$ is a subset of $V(G)$.
We call a cut $\delta(S)$ \textit{non-trivial} if $S$ is a nonempty,
proper subset of $V$, that is, if $\emptyset\neq{S}\subsetneq{V}$.

The following integer program formulates the Cap-$k$-ECSS problem.
It can be viewed as the natural ``cut covering'' formulation of the problem.
It has a binary variable $x_e$ for each edge $e$, with the meaning
that an edge $e$ is picked iff $x_e=1$, and, for each non-trivial
cut, it has a constraint stating that the capacity of the picked
edges in the cut is $\geq{k}$.
\begin{align*}\label{intro-IP:CapkECSS}
    \min\;\;&\sum_{e\in E}c_e x_e \tag{IP: CapkECSS}\\
    \text{s.t.}\;\;& \sum_{e\in E\cap \delta(S)} u_ex_e \geq k && \forall\;\; \emptyset\subsetneq S\subsetneq V \\
    &x_e\in \{0,1\} && \forall\;\; e\in{E}
\end{align*}
The LP (linear programming) relaxation of the above integer program
is obtained by replacing $x_e\in\{0,1\}$ by $0\leq x_e\leq 1$,
$\forall e\in{E}$.
The following well-known example shows that the LP relaxation has
an integrality ratio of $\Omega(k)$;
similar examples are given in \cite{CFLP00,CCKK15}.

\begin{example} \label{example:IR-CapkECSS}
The graph $G$ consists of two nodes $u,v$, and a pair of parallel
edges $e_1, e_2$ between the two nodes.  Edge $e_1$ has cost zero
and capacity $k-1$, and edge $e_2$ has cost~one and capacity $k$.
A feasible solution of the integer program has cost one since the
edge $e_2$ must be chosen in any feasible solution.
On the other hand, a feasible solution $x$ to the LP relaxation of
cost $1/k$ is given by $x_{e_1}=1, x_{e_2}=1/k$.
Hence, the integrality ratio of the LP relaxation is $k$ for this example.
Thus, we face an obstruction for the task of designing any approximation
algorithm that achieves approximation ratio $o(k)$ by rounding this LP relaxation.
\end{example}
}

\subsection{Flexible Graph Connectivity and the \texorpdfstring{$\pqfgc$}{(p,q)-FGC} problem}
{
Adjiashvili, Hommelsheim and M\"uhlenthaler \cite{AHM22}
introduced the model of Flexible Graph Connectivity that we denote by $\fgc$ as a way to model network design problems where edges have non-uniform reliability. 
Boyd, Cheriyan, Haddadan and Ibrahimpur \cite{BCHI24} introduced a
generalization of $\fgc$, called the $(p,q)$-Flexible Graph Connectivity problem, denoted $\pqfgc$, where $p$ is
an integer denoting the {connectivity} requirement, and $q$ is an integer denoting the {robustness} requirement.
An instance of $\pqfgc$ consists of an undirected graph $G = (V,E)$, where $E$ is partitioned into
a set of safe edges $\safe$ (edges that never fail) and a set of unsafe edges $\unsafe$ (edges that may fail),
and nonnegative edge-costs $c \in \Qp^E$.  A subset $F \subseteq
E$ of edges is feasible for the $\pqfgc$ problem if for any set
$F'$ consisting of at most $q$ unsafe edges, the subgraph $(V, F-F')$
remains $p$-edge connected. The objective is to find a
feasible solution $F$ that minimizes $c(F) = \sum_{e \in F} c_e$.

The following linear program gives a lower bound on the optimal value for $(p,q)$-FGC.
Such LP relaxations are discussed in \cite{BCHI24} and \cite[Section~2]{CJ23}.
To motivate the LP relaxation, consider an auxiliary capacitated graph
that has the same set of nodes and the same set of edges as the
graph of the $(p,q)$-FGC instance.
Assign a capacity of $(p+q)$ to each safe edge and a capacity of $p$ to each unsafe edge.
Let $k=p(p+q)$ and view the capacitated graph as an instance of the Cap-$k$-ECSS problem.
In general, observe that a feasible solution of the $(p,q)$-FGC instance corresponds
to a feasible solution of the Cap-$k$-ECSS instance, but not vice-versa.
(When either $p=1$ or $q=1$, then a feasible solution of the Cap-$k$-ECSS instance
corresponds to a feasible solution of the $(p,q)$-FGC instance.)
Each edge $e\in\safe\cup\unsafe$ has a variable $x_e$.
\begin{align}
	\min  & \sum_{e \in \safe\cup\unsafe} c_e x_e & \label{eq:obj} \\
	\text{s.t.} \quad & \sum_{e\in\safe\cap\delta(S)} (p+q)\; x_e ~~+~~ \sum_{e\in\unsafe\cap\delta(S)} (p)\; x_e \geq p(p+q)  & \forall \quad \emptyset \subsetneq S \subsetneq V \label{eq:twoec} \\
& 0 \leq x_e \leq 1 & \forall e\in\safe\cup\unsafe	\label{eq:nonneg}
\end{align}
Unfortunately, similarly to the LP relaxation for Cap-$k$-ECSS, the
above LP relaxation has a large integrality ratio, even for the
special case of $\oneqfgc$. Example~\ref{example:IR-CapkECSS} can be modified such that for $p=1$ and $q>0$, the above
LP relaxation has integrality ratio $(q+1)$.
}

\subsection{Our results}
{
Our first result is an $O(\log(k/u_{min}))$ approximation algorithm
for the Cap-$k$-ECSS problem.
Thus, we asymptotically improve upon the previous best approximation ratio of
  $\min(O(\log{n}),\; k,\; 2u_{max},\; \pliableapxOLD \cdot {\lceil k/u_{min} \rceil})$
whenever $\log(k/u_{min})=o(\log{n})$ and $u_{max}$ is sufficiently large.

\begin{restatable}{theorem}{CapkECSStheorem}
\label{thm:approx-CapkECSS}
There is a polynomial-time algorithm that, given an instance of Cap-$k$-ECSS, computes a vector $x^*$ of cost at most $\opt$ that possibly satisfies only a subset of the constraints of
\eqref{intro-KCLP:CapkECSS}, and rounds it to a feasible integer solution of cost at most $O(\log(k/u_{min})) \cdot \opt$.
\end{restatable}

Recall that the $\oneqfgc$ problem can be formulated as a special
case of Cap-$k$-ECSS with two capacities (the unsafe edges have
unit capacity and the safe edges have capacity $k=(q+1)$).
Our next result improves the approximation ratio to $\oneqfgcapx$, thus improving on
the previous best $(q+1)$-approximation algorithm for $\oneqfgc$, \cite{BCHI24}.

\begin{restatable}{theorem}{oneqFGCtheorem}
\label{thm:approx-oneqfgc}
There is a polynomial-time algorithm that, given an instance of $\oneqfgc$, computes a vector $x^*$ of cost at most $\opt$ that possibly satisfies only a subset of the constraints of \eqref{KCLP:oneqfgc}, and rounds it to a feasible integer solution of cost at most $\oneqfgcapx\,\opt$.
\end{restatable}

Moreover, we present $O(1)$-approximate reductions between the
$\twoqfgc$ problem and the $\twoASC$ problem. The following two
results summarize the two reductions.

\begin{restatable}{theorem}{twoqFGCreduction}
\label{thm:approx-twoqfgc}
Suppose an LP~relative $\rho$-approximation algorithm
for $\twoASC$ that runs in polynomial~time is available (assume $\rho\geq1$).
Then there is an algorithm for $\twoqfgc$ that runs in polynomial
time and returns a feasible (integer) solution of cost $\leq (4(\rho+1)+\oneqfgcapx)\,\opt$.
\end{restatable}

\begin{restatable}{theorem}{twoASCreduction}
\label{thm:approx-twoASC}
Suppose a $\rho'$-approximation algorithm
for $\twoqfgc$ that runs in polynomial~time is available.
Then there is an algorithm for unit-capacity $\twoASC$ that runs in polynomial
time and returns a feasible (integer) solution of cost $\leq (\rho'+2)\,\opt$.
\end{restatable}

}

\subsection{Organization of the Paper}

For the sake of readability and accessibility, we present our
$\oneqfgcapx$-approximation algorithm for $\oneqfgc$ in
section~\ref{sec:oneqfgc}, and we defer the presentation of our
improved approximation algorithm for Cap-$k$-ECSS to
section~\ref{sec:logk-approx}.
The $O(1)$-approximate reductions between $\twoqfgc$ and $\twoASC$
are presented in section~\ref{sec:twoqfgc}.
}
%%%%%%%%%%
{
\section{Preliminaries:
	Techniques, Tools, and Overview of Approximation Algorithms \label{sec:prelims}}

In this subsection, we describe three of the known tools that we apply
\textit{together} to obtain our main results.  Each of these
tools has been used on its own (without the other tools) to obtain
improvements in the approximation ratio of the Cap-$k$-ECSS problem,
but, in our opinion, by combining these tools in the right way, we
obtain striking improvements in the approximation ratios for the
Cap-$k$-ECSS problem and the $\oneqfgc$ problem.

The first tool is the algorithmic use of the Knapsack-Cover
Inequalities (KCI) for strengthening the LP relaxation of \eqref{intro-IP:CapkECSS}.
This tool was introduced by Carr, Fleischer, Leung \& Phillips \cite{CFLP00}.
The second tool is an $O(1)$~approximation algorithm for the so-called $\ASC$ problem.
This tool was introduced by Bansal, Cheriyan, Grout \& Ibrahimpur \cite{BCGI24}.
The third tool is Jain's iterative rounding method, \cite{Jain01}.

\subsection{Knapsack-Cover Inequalities (KCI) for Capacitated Network Design \label{sec:KCI}}
{
Our approximation algorithms for both Cap-$k$-ECSS and
$\oneqfgc$ use the LP relaxations highlighted
above as the starting point. However, to eliminate the integrality gap,
we will strengthen these relaxations using knapsack-cover
inequalities. We focus here on illustrating this tool for the
Cap-$k$-ECSS problem, strengthening \eqref{intro-IP:CapkECSS}.

For any non-trivial cut $\delta(S)$ and a subset of the edges
$A\subseteq E$, the following is a valid inequality for all integer
solutions of \eqref{intro-IP:CapkECSS}.
\[
        \sum_{e\in E\cap\delta(S) - A} u_e(A,S) x_e \geq D(A,S),
\]
where $D(A,S) = \max\{0,\, k - \sum_{e\in \delta(S)\cap A} u_e\}$ and
$u_e(A,S) = \min\{u_e,\, D(A,S)\}$.
(By plugging in $A=\emptyset$, we get the constraint
$\sum_{e\in E\cap\delta(S)} u_e x_e \geq k$,
which is a constraint of \eqref{intro-IP:CapkECSS}.)
Intuitively, the added knapsack-cover inequalities for a cut
$\delta(S)$ and edge-set $A$ ensure that if a high-capacity edge
is being used to cover $\delta(S)$ and, moreover, $\delta(S)$ is
covered by some of the edges in $A$, then the capacity of the high-capacity
edge is reduced to the remaining requirement, namely, $k-u(A\cap\delta(S))$.
These inequalities ``cut off'' poor solutions $x$ of the original LP~relaxation
(i.e., the one without KCI) such that some high-capacity edge $f$ has
a small fractional value for $x_f$ (i.e., $0<x_f\ll{1}$).
In particular, for Example~\ref{example:IR-CapkECSS} (at the end of section~\ref{sec:intro:cnd}),
consider the knapsack-cover inequality for the cut $\delta(S)$ where
$S=\{u\}$ and $A=\{e_1\}$. We have $D(A,S)=\max\{0,\, k-(k-1)\}=1$
and $u_{e_2}(A,S)=\min\{u_{e_2},\, D(A,S)\}=\min\{k,1\}=1$, hence, this inequality is
$\sum_{e\in\delta(S)-A} u_e(A,S)\, x_e \geq D(A,S)$ which is
$ x_{e_2} \geq 1$.
Clearly, the fractional solution $x_{e_1}=1, x_{e_2}=1/k$  is ``cut off'' by this inequality.

We add these inequalities to \eqref{intro-IP:CapkECSS} to obtain the
following LP relaxation of the Cap-$k$-ECSS problem.
\begin{align*}\label{intro-KCLP:CapkECSS}
    \min\;\;&\sum_{e\in E}c_e x_e \tag{KCLP:~CapkECSS}\\
    \text{s.t.}\;\;& \sum_{e\in E\cap\delta(S) - A} u_e(A,S) x_e \geq D(A,S) && \forall\;\; \emptyset\subsetneq S\subsetneq V, A\subseteq E \\
    &0 \leq x_e \leq 1 && \forall\;\; e\in E
\end{align*}
Observe that this LP has a number of constraints that is exponential
in the size of the input~instance (of Cap-$k$-ECSS).
Moreover, we do not know of any polynomial-time separation
oracle for the entire set of knapsack-cover inequalities.
Nevertheless, by following the cut-and-round approach employed by
Carr, Fleischer, Leung \& Phillips \cite{CFLP00},
one can round this LP in polynomial time to an approximately optimal
integer solution via the ellipsoid~method by designing efficient subroutines.
See sections~\ref{sec:oneqfgc}, \ref{sec:logk-approx},
	for details.

{
Recently, Ibrahimpur \& V\'{e}gh \cite{IV:ipco2025} presented a polynomial-time
separation subroutine for the knapsack-cover inequalities for the $\pqfgc$~problem.
}
}

\subsection{The \texorpdfstring{$\ASC$}{Cover Small Cuts} problem}
{
We follow the notation from \cite[Section~1.3]{BCGI24}.
In an instance of the $\ASC$ problem, we are given an
undirected capacitated graph $G = (V,E)$ with edge-capacities $u
\in \Qp^E$, a set of links $L \subseteq \binom{V}{2}$ with costs
$c \in \Qp^L$, and a threshold $\capbound \in \Qp$.
A subset $F \subseteq L$ of links is said to \emph{cover} a
node-set $S$ if there exists a link $e \in F$ with exactly one
end-node in $S$.
The objective is to find a minimum-cost $F\subseteq{L}$ that covers
each non-empty $S \subsetneq V$ with $u(\delta_E(S)) < \capbound$.
Let $\C = \{ \emptyset \neq S \subsetneq V : u(\delta_E(S)) < \capbound \}$.
Then we have the following covering~LP relaxation of the problem.
\begin{align*}\label{LP:ASC} \tag{LP:\,{Cover\,Small\,Cuts}}
\min \quad \qquad		& \quad \sum_{f \in L} c_f x_f 		& \\
\text{subject to: } 	& \quad \sum_{f\in L\cap\delta(S)} x_f \geq 1 	& \forall \, \, S \in \C \\
					& \quad 0\leq x_f \leq 1 & \forall \, \, f \in L \\
\end{align*}
The following result is due to Simmons et~al.~\cite{S2025,SBC-5approx};
for previous results on $\ASC$, see Bansal \cite{B2023,B:ipco2025} and Nutov \cite{N:mfcs2025}.
\begin{proposition} \label{prop:approxASC}
Given an instance of $\ASC$, the WGMV primal-dual algorithm, \cite{WGMV95}, finds
a feasible solution of cost $\leq \pliableapx\,\lpopt$ in polynomial time,
where $\lpopt$ denotes the optimal value of \eqref{LP:ASC}.
\end{proposition}

In the $\twoASC$ problem,
the inputs are the same as above, namely, $G=(V,E), u, L, c, \capbound$.
A subset $F \subseteq L$ of links is said to \emph{two-cover} a
node-set $S$ if $|F\cap\delta(S)|\geq2$, that is,
if there exist a pair of (distinct) links $e,e' \in F$
such that each of $e$ and $e'$ has exactly one end-node in $S$.
The objective is to find a minimum-cost $F\subseteq{L}$ that two-covers
each non-empty $S \subsetneq V$ with $u(\delta_E(S)) < \capbound$.
}

\subsection{The \texorpdfstring{$f$-connectivity}{f-connectivity} problem and Jain's iterative rounding algorithm \label{sec:f-conn}}
{
In the context of approximation algorithms, several connectivity
augmentation problems can be formulated in a general framework
called $f$-connectivity. In this problem, we are given an undirected
graph $G = (V,E)$ on $n$ nodes with nonnegative costs $c \in \Qp^E$
on the edges and a requirement function $f:2^V\to\Zint$ on subsets of nodes.
The algorithmic goal is to find an edge-set $\J \subseteq E$ with
minimum cost $c(\J) := \sum_{e \in \J} c_e$ such that for all cuts
$\delta(S),\ S \subseteq V$, we have $|\delta(S) \cap \J| \geq f(S)$.
{
A function $f$ is called \textit{weakly supermodular} if
$f(V)=0$, and
for all $A,B \subseteq{V}$, either
    $f(A) + f(B) \leq f(A - B) + f(B - A)$, or
    $f(A) + f(B) \leq f(A \cap B) + f(A \cup B)$.
}

Assuming that the function $f$ is weakly supermodular, {integral,
and has a positive value for some $S\subset{V}$}, Jain \cite{Jain01}
presented a 2-approximation algorithm for the $f$-connectivity problem.

We will apply Jain's result (or its extension) in most of our algorithms.
}

\subsection{Overview of Our Approximation Algorithms}
{
First, we focus on the special case of Cap-$k$-ECSS with capacities $1$ and $k$.
Let $E^{(1)}$ be the set of unit-capacity edges, and let $E^{(k)}$ be the set of edges of capacity $k$.
For expository reasons (and glossing over some critical points),
let us assume that we can find, in polynomial time, an optimal
solution $x^*$ of \eqref{intro-KCLP:CapkECSS}, the LP with KCI.
Thus, we have $c(x^*)=\lpopt$, where $c(x^*)=\sum_e c_ex_e$ is the cost of $x^*$.
Let $E^{(1)}_{high} = \{e\in E^{(1)} : x^*_e \geq 1/2\}$;
this is the set of unit-capacity edges that are assigned values of $1/2$ or more by the LP solution $x^*$.
Let $E^{(1)}_{low} = E^{(1)} - E^{(1)}_{high}$ (the set of unit-capacity edges with $x^*$-values less than $1/2$).
We call a non-trivial cut $\delta(S)$ a {small cut} if
\[
|E^{(1)}_{high}\cap\delta(S)| \;+\;
\sum_{e\in{E^{(1)}_{low}\cap\delta(S)}}{2} x^*_e < k; \qquad \tag{definition of small cut}
\]
in other words, a non-trivial cut is defined to be small if
the fractional capacity contributed by the unit-capacity edges is less than the requirement of $k$
even after rounding up edges in $E^{(1)}_{high}$ to have $x$-value one
and scaling up the $x$-value of each edge in $E^{(1)}_{low}$ by factor $2$.
We construct an instance of $\ASC$ with small cuts as defined above,
and we define the link-set of this instance to be $E^{(k)}$.
To handle the small cuts, we apply the knapsack-cover inequalities
to show that the solution $x^*$ restricted to edges of capacity~$k$ and scaled up by a factor of $2$ (i.e., $2 x^*_{E^{(k)}}$)
constitutes a feasible solution to the LP relaxation of the $\ASC$ instance.
This is the critical point where our algorithm and analysis relies on the knapsack-cover inequalities.
We can thus pick a set $E^{(k)}_{picked}$ of edges of capacity~$k$ by applying the
$\pliableapx$-approximation algorithm of \cite{S2025,SBC-5approx} to this instance
of the $\ASC$ problem;
note that the cost of $E^{(k)}_{picked}$ is at most $\pliableapx$ times the cost of $2 x^*_{E^{(k)}}$.
After this step,
we contract the connected components formed by the edges in $E^{(k)}_{picked}$,
and get a new instance that does not have any small cuts.
Since there are no small cuts, every non-trivial cut $\delta(S)$
satisfies the inequality
$|E^{(1)}_{high}\cap\delta(S)| \;+\;
\sum_{e\in{E^{(1)}_{low}\cap\delta(S)}} {2} x^*_e \geq k$.
Therefore, we get a feasible solution for the LP relaxation of the $f$-connectivity
problem where $f(S) = k$ for every non-empty set $S\subsetneq{V}$,
by picking all edges in $E^{(1)}_{high}$ and scaling up $x^*$ restricted
to $E^{(1)}_{low}$ by a factor of $2$.
Since $f$ is weakly supermodular, we can apply Jain's iterative
rounding method~\cite{Jain01} to solve this $f$-connectivity problem
and obtain a $2$-approximate solution, giving us an integral
solution whose cost is at most $4$ {times} the cost of
$x^*$ restricted to unit-capacity edges.
Thus, we get an integral solution whose total cost 
is at most $\pliableapx\cdot{2}\cdot\lpopt$.

Let us recap the new algorithmic lever we deployed above.
We partitioned the edges according to their capacities into the set
of low-capacity edges and the set of high-capacity edges; let
$E_{big}$ denote the latter set.
Then we defined an instance of $\ASC$ whose small cuts were defined
using the low-capacity edges and rounding up (and/or scaling up)
the fractional capacity contribution $u_ex^*_e$ of each of these
edges $e$; moreover, we defined the links (of the $\ASC$ instance)
to be the high-capacity edges.
The small cuts identified above are precisely the cuts where the
LP solution $x^*$ has not invested sufficient fractional capacity
in the low-capacity edges to cover the cuts. For these cuts, the
knapsack-cover inequalities ensured that $x^*$ restricted to the
high-capacity edges and scaled up by a small constant, say, $\eta$,
(i.e., $\eta x^*_{E_{big}}$) forms a feasible solution to the LP
relaxation of the $\ASC$ instance.
Based on this, we applied the $\pliableapx$-approximation algorithm
of \cite{S2025,SBC-5approx} to find a set of high-capacity edges
of cost $\leq \pliableapx\,\eta c(x^*_{E_{big}})$ that covers all the small cuts.

Now, let us sketch an extension of the above method that gives
an $O(\log{k})$-approximation algorithm for Cap-$k$-ECSS.
As above, let us assume that we can find, in polynomial time, an optimal
solution $x^*$ of \eqref{intro-KCLP:CapkECSS}, the LP with KCI.
Thus, we have $c(x^*)=\lpopt$.
Throughout the execution of the algorithm, we maintain a set of
edges $E_{cur}$ acting as our current solution.
We begin with $E_{cur} = \{e\in E:x^*_e \geq 1/2\}$.
Define $T = \lceil\log{k}\rceil$ and
for $j=1,2,\ldots,T$,
define $E_j = \{e\in E: x^*_e < 1/2 \text{ and } u_e \leq 2^j\}$.
Ideally, we wish to apply $T$ iterations, and, in each iteration, we wish to apply the above algorithmic lever $O(1)$ times.
In more detail, in the $i$th-iteration, we could take the edges of capacity $\leq 2^{T-i}$ to be the low-capacity edges and the edges of capacity $>2^{T-i}$ to be the high-capacity edges; that is, $E_{T-i}$ is the set of low-capacity edges and $E-E_{T-i}$ is the set of high-capacity edges. Then, we could define an instance of $\ASC$ where we could define the small-cuts to be the non-trivial cuts $\delta(S)$ such that
$(\sum_{e\in{E_{cur}}\cap\delta(S)} u_e) \;+\;
(\sum_{e\in{E_{T-i}}\cap\delta(S)}{2} u_e x^*_e) < k$,
and we could take the link-set (for the $\ASC$ instance) to be the set of high-capacity edges, $E-E_{T-i}$.
Although we can apply the algorithmic lever and find an integral cover of the small cuts using the edges in $E-E_{T-i}$ such that the cost of the integral cover is $O(1) c(x^*_{E-E_{T-i}})$, we run into a difficulty.
The capacity of an edge in $E-E_{T-i}$ could be as small as $\Theta(2^{T-i})$, hence, by adding just one of these edges to a small cut $\delta(S)$ we cannot guarantee that the capacity of $\delta(S)$ is augmented to become $\geq k$.
Thus, our algorithm and analysis, presented in section~\ref{sec:logk-approx}, have further steps; a deeper analysis is required to show that our algorithm finds an integral solution of cost $O(\log{k}) \cdot \lpopt$.
Moreover, we improve the approximation ratio from $O(\log{k})$ to $O(\log(k/u_{min}))$.
}
}
%%%%%%%%%%
{
\section{\texorpdfstring{$\oneqfgcapx$}{7}-Approximation Algorithm for \texorpdfstring{$\oneqfgc$}{(1,q)-FGC} \label{sec:oneqfgc}}

This section presents a $\oneqfgcapx$-approximation algorithm for the $(1,q)$-FGC problem.
For the convenience of the reader, the presentation in this section
is independent of the rest of the paper.
For a graph $G=(V,E)$ and $S\subseteq{V}$, the cut $\delta(S)$
refers to the set of edges that have exactly one end-node in $S$;
$\delta(S)$ is called a {non-trivial} cut if $\emptyset\neq{S}\subsetneq{V}$.
(Whenever we use the term ``cut $\delta(S)$'' we mean that $S$ is a subset of $V(G)$.)
We use the term \textit{small cut} to mean a non-trivial cut
$\delta(S)$ with capacity below a specified threshold-value, say, $\lambda$.

Our starting point is the natural LP relaxation that follows from
taking a capacitated network design view of the problem where each
unsafe edge $e \in \unsafe$ has capacity $u_e=1$, and
each safe edge $e \in \safe$ has capacity $u_e=(q+1)$. The natural
LP relaxation then seeks to minimize the total cost of edges subject
to the constraint that $\forall\;\; \emptyset\subsetneq S\subsetneq
V$, we have $\sum_{e\in\safe\cap\delta(S)} (q+1) x_e +
\sum_{e\in\unsafe\cap\delta(S)} x_e \geq (q+1).$

It is easy to see that any feasible solution to this LP with zero/one
values is a valid solution to a given instance of  $\oneqfgc$, and
vice versa.
However, it is also easy to show that this LP has integrality ratio
$(q+1)$ by adapting Example~\ref{example:IR-CapkECSS} given
at the end of section~\ref{sec:intro:cnd}.

To get around this obstruction, we strengthen the LP relaxation of
$\oneqfgc$ using the knapsack-cover inequalities to obtain the
following stronger LP. Intuitively, the added knapsack-cover
inequalities for a cut $\delta(S)$ ensure that if a safe edge is
being used to cover $\delta(S)$ and, moreover, $\delta(S)$ is partly
covered by unsafe edges, say, by $\ell$ of them, then the capacity
of the safe edge is reduced to $(q+1)-\ell$.
\begin{align*}\label{KCLP:oneqfgc}
    \min\;\;&\sum_{e\in E}c_e x_e \tag{KCLP:$(1,q)$-FGC}\\
    \text{s.t.}\;\;& \sum_{e\in\safe\cap\delta(S)} (q+1) x_e + \sum_{e\in\unsafe\cap\delta(S)} x_e \geq (q+1) && \forall\;\; \emptyset\subsetneq S\subsetneq V \\
    & \sum_{e\in E\cap\delta(S) - A} u_e(A,S) x_e \geq D(A,S) && \forall\;\; \emptyset\subsetneq S\subsetneq V, A\subseteq E \\
    &0 \leq x_e \leq 1 && \forall\;\; e\in E,
\end{align*}
where $D(A,S) = \max\{0,\, (q+1) - \sum_{e\in \delta(S)\cap A} u_e\}$, and
$u_e(A,S) = \min\{u_e,\, D(A,S)\}$.
{
We mentioned above that Ibrahimpur \& Vegh \cite{IV:ipco2025} have presented a
polynomial-time separation subroutine for the knapsack-cover inequalities
for the $\pqfgc$~problem.  Instead of using their method, we follow
the algorithmic scheme of \cite{CFLP00,CCKK15}, because we will use a
similar algorithmic scheme in section~\ref{sec:logk-approx} (to the best
of our knowledge, there is no polynomial-time separation subroutine for
the knapsack-cover inequalities in section~\ref{sec:logk-approx}).
Our plan is to identify a subset 
of unsafe edges $A$ and a collection of sets $\C$ (in polynomial time)
such that, as long as the knapsack-cover inequalities hold for $A$ and all $S\in\C$,
we will be able to execute our rounding algorithm.
Using this, we will design a polynomial-time approximation algorithm for $\oneqfgc$.

In this section, let $\alpha$ denote the best approximation ratio known for the $\ASC$ problem.
We will design an $(\alpha+2)$-approximation algorithm for the $\oneqfgc$ problem.
As of now, we have $(\alpha+2)=\oneqfgcapx$, since $\alpha=\pliableapx$,
due to \cite{S2025,SBC-5approx}, see Proposition~\ref{prop:approxASC}.
}

{Let $\lpopt$ denote the optimal value of \eqref{KCLP:oneqfgc},
the LP with the knapsack-cover inequalities.  Using binary search
for $\lpopt$ together with the ellipsoid algorithm and our
polynomial-time subroutines, we will find a vector $x^*$ with
cost $c(x^*)=\sum_e{c_ex_e}$ such that $c(x^*) \leq
\lpopt$ and $x^*$ satisfies the constraints of \eqref{KCLP:oneqfgc}
specified in the next lemma (though $x^*$ could violate other
constraints of \eqref{KCLP:oneqfgc}). Then, we will ``round'' $x^*$
to an integer solution of cost $\leq (\alpha+2) c(x^*)$. Thus,
we will find an integer solution of cost $\leq (\alpha+2) \lpopt$
(even though we will not compute the precise value of $\lpopt$).
We discuss our overall algorithm and the outer loop of binary search
at the end of this section.
}

\begin{lemma}\label{lem:polytimeKC-oneqfgc}
There is a polynomial-time algorithm that given a vector $x^*$ 
(that is a candidate solution of \eqref{KCLP:oneqfgc}) and
a value $z$
either finds a violated constraint of the LP
or else verifies that $c(x^*)\leq{z}$ and, moreover, $x^*$ satisfies the following two properties:
\begin{itemize}
\item[(P1)]
$\sum_{e\in\safe\cap\delta(S)} (q+1) x^*_e + \sum_{e\in\unsafe\cap\delta(S)} x^*_e \geq (q+1) \qquad
  \forall\: \emptyset\neq{S}\subsetneq{V}$.

\item[(P2)]
	Let $\ds A = \{ e \in  \unsafe~|~x^*_e \ge \frac{2}{(\alpha+2)} \}.$
For any non-empty ${S}\subsetneq{V}$,
if $\sum_{e\in\safe\cap\delta(S)} (q+1) x^*_e +
\sum_{e\in\unsafe\cap\delta(S)} x^*_e \leq 2(q+1)$, then
$\ds \sum_{e\in E\cap\delta(S) - A} u_e(A,S) x^*_e \geq D(A,S).$
\end{itemize}
\end{lemma}
\begin{proof}
Essentially, we will describe a polynomial-time separation oracle that 
identifies any violations of properties~(P1) and~(P2).
Given a vector $x^*$, we first check that $\sum_{e \in E} c_e x_e^* \le z$. 
If not, we return this as a violated constraint. Otherwise,
let $\hG=(\hV,\hE)$ be the capacitated graph where $\hV=V,\hE=E,$
and each edge $e \in \hE$ is assigned a capacity of $u_ex_e^*$. We
can now check that the capacity of a minimum-cut of $\hG$ is at
least $(q+1)$ using a polynomial-time global minimum-cut algorithm
\cite{Schrijver-book}.  If not, we return a global minimum cut in
$\hG$ as a violated constraint. Otherwise, we know that (P1) is
satisfied, and we proceed to verify (P2) with respect to the set
$\ds A = \{ e \in  \unsafe~|~x^*_e \ge \frac{2}{(\alpha+2)} \}$.

By Karger's result \cite{Karger93}, there are at most $O(n^4)$ cuts
of capacity at most $2(q+1)$ (i.e., at most twice the capacity of
a minimum~cut), and, moreover, we can enumerate all such cuts of
$\hG$ in polynomial time \cite{NNI97}.
By iterating over each of the $O(n^4)$ cuts, we can now verify, in
polynomial~time, that the knapsack-cover inequalities are satisfied
w.r.t.\ the set $A$. If not, we have found a violated constraint.
\end{proof}

The next corollary follows from the above lemma and the well-known fact that 
the ellipsoid algorithm terminates after {$n^{O(1)}$} iterations
of feasibility verification \cite{ellipsoid-book}.
The outer loop of our algorithm runs a binary search for $\lpopt$ 
(see the discussion at the end of this section).

\begin{corollary}
There is a polynomial-time algorithm that computes a vector $x^*$ of cost
$c(x^*)\leq\lpopt+\epsilon$ such that $x^*$ satisfies properties~(P1) and~(P2),
	for any desired $\epsilon>0$.
Possibly, $x^*$ violates some of the other constraints of \eqref{KCLP:oneqfgc}.
\end{corollary}

\smallskip

\noindent
{\bf The Rounding Algorithm:}
Given a value $z$ and a vector $x^*$ of cost $c(x^*)\leq{z}$ that satisfies
properties~(P1) and~(P2) (see Lemma~\ref{lem:polytimeKC-oneqfgc}),
Algorithm~\ref{alg:oneq-FGC}, presented below, rounds it to an
integer solution of cost at most $(\alpha+2) {z}$. 
Below, we describe the main idea of our rounding scheme.

Recall that our LP relaxation assigns capacity $u_e=1$ for each
unsafe edge $e\in\unsafe$ and capacity $u_e=(q+1)$ to each safe edge $e\in\safe$.
We call a non-trivial cut $\delta(S)$ a {small cut} if
$|\unsafe_1\cap\delta(S)| \;+\;
\sum_{e\in\unsafe_2\cap\delta(S)}\frac{(\alpha+2)}{2} x^*_e < (q+1)$
where $\ds \unsafe_{1} = \{e\in \unsafe : x^*_e \geq \frac{2}{(\alpha+2)} \}$, and
$\unsafe_{2} = \unsafe - \unsafe_1$.
To handle the presence of small cuts, 
we first construct an instance of $\ASC$ with the specified small cuts
and with link-set the set of safe edges;
then, via Lemma~\ref{lem:oneqfgc1}, we show that the vector $x^*$ restricted to safe edges and 
scaled up by a factor of $\ds \frac{(\alpha+2)}{\alpha}$ (i.e., $\frac{(\alpha+2)}{\alpha} x^*_{\safe}$)
constitutes a feasible solution to the LP relaxation of the $\ASC$ instance.
We can thus pick a set $\safe_1$ of safe edges by applying the
$\alpha$-approximation algorithm to the $\ASC$ instance; 
note that the cost of $\safe_1$ is at most $\alpha$ times the cost of $\frac{(\alpha+2)}{\alpha} x^*_{\safe}$.
After this step, we contract the connected components formed by the edges in $\safe_1$,
and get a new instance that does not have any small cuts.
Since there are no small cuts, every non-trivial cut $\delta(S)$
satisfies the inequality
$\ds |\unsafe_1\cap\delta(S)| \;+\;
\sum_{e\in\unsafe_2\cap\delta(S)}\frac{(\alpha+2)}{2} x^*_e \geq (q+1)$.
Therefore, we get a feasible solution for the LP relaxation of the 
$f$-connectivity problem where $f(S) = q+1$ for every non-empty set $S\subsetneq{V}$,
by picking all edges in $\unsafe_1$ and scaling up $x^*$ restricted
to $\unsafe_2$ by a factor of $(\alpha+2)/2$.
Since $f$ is weakly supermodular, we can apply Jain's iterative
rounding method~\cite{Jain01} to solve this $f$-connectivity problem
and obtain a $2$-approximate solution, giving us an integral
solution whose cost is at most $(\alpha+2)$ {times} the cost of
$x^*_{\unsafe}$ (see Lemma~\ref{lem:oneqfgc2}).
Thus, we get an integral solution whose total cost of safe and
unsafe edges is at most $(\alpha+2)$ times $\lpopt$.

%-----
{
 \begin{algorithm}
     \caption{$(\alpha+2)$-approximate solution to $(1,{q})$-FGC} \label{alg:oneq-FGC}
      \textbf{Require:} Graph $G=(V,E)$ where $E$ is partitioned into $\safe\bigcup\unsafe$, with edge costs $\{c_e\}_{e\in E}$.  A solution $x^*$ to \eqref{KCLP:oneqfgc} as promised by Lemma~\ref{lem:polytimeKC-oneqfgc}.
     \begin{enumerate}
	     \item $\ds \unsafe_{1} \gets \{e\in \unsafe : x^*_e \geq \frac{2}{(\alpha+2)} \}$, and $\unsafe_{2} \gets \unsafe - \unsafe_1$.
         \item
		{\begin{enumerate}
             \item[\mbox{~~~~}(a)]
		Apply the approximation~algorithm for $\ASC$ on the instance with
		$G'=(V,E' = \unsafe)$,
		where each edge of $\unsafe_1$ is given unit capacity,
		each edge $e\in\unsafe_2$ is given capacity $\ds \frac{(\alpha+2)}{2} x^*_e$,
		and with link set $\safe$.
		Define the threshold $\lambda$ (for small~cuts) to be $(q+1)$. Thus, we have:
\[
	|\unsafe_1\cap\delta(S)| \;+\; \sum_{e\in \unsafe_2\cap\delta(S)}\frac{(\alpha+2)}{2} x^*_e < (q+1) \tag{definition of {\em small cuts}}
\]
             \item[\mbox{~~~~}(b)] 
		Let the output of the call in step~(a) be denoted $\safe_{1}$.
		\end{enumerate}}
         \item Construct a graph $G''=(V,\unsafe_{2})$ where each edge $e\in \unsafe_{2}$ has cost $c_e$.
		Define the requirement of a non-trivial cut $\delta(S)$ to be
$$f(S) = (q+1) - \big((q+1) |\delta(S)\cap \safe_{1}| + |\delta(S)\cap \unsafe_{1}|\big).$$
		This function $f$ is weakly supermodular (note that $f(V)=f(\emptyset)=0$), so
		a 2-approximate solution for this instance of $f$-connectivity can be
		computed using Jain's iterative rounding method \cite{Jain01}.
         \item Return the union of the set of unsafe~edges picked by the previous step (via the iterative rounding algorithm) and $\safe_{1} \cup \unsafe_{1}$.
     \end{enumerate}
 \end{algorithm}
}
%-----

The next two lemmas formalize the key properties of the solution
$x^*$ that are used in the rounding scheme above, allowing us to
show that it returns a feasible integral solution of cost at most
$(\alpha+2) c(x^*)$.

\begin{lemma}\label{lem:oneqfgc1}
In step~2 of Algorithm~\ref{alg:oneq-FGC}, a feasible fractional
solution to the $\ASC$ instance is given by
	$\ds \hat{x}_e = \min\{1,\, \frac{(\alpha+2)}{\alpha} x^*_e\}$ for $e\in \safe$.
\end{lemma}
\begin{proof}
Consider any small cut $\delta(S)$. We will establish the lemma by considering two cases.

First, consider the case that $\sum_{e\in\safe\cap\delta(S)} (q+1)
x^*_e + \sum_{e\in\unsafe\cap\delta(S)} x^*_e > 2(q+1)$. Since
$\delta(S)$ is a small cut, we have
\[
\sum_{e\in \unsafe_1\cap\delta(S)}x^*_e \;+\; \sum_{e\in \unsafe_2\cap\delta(S)}x^*_e \leq
|\unsafe_1\cap\delta(S)| \;+\; \sum_{e\in \unsafe_2\cap\delta(S)}\frac{(\alpha+2)}{2} x^*_e <
	(q+1).
\]
Since $\sum_{e\in\safe\cap\delta(S)} (q+1) x^*_e +
\sum_{e\in\unsafe\cap\delta(S)} x^*_e > 2(q+1)$, it follows that
$\sum_{e\in\safe\cap\delta(S)} (q+1) x^*_e \geq (q+1)$, hence,
$\sum_{e\in\safe\cap\delta(S)} x^*_e \geq 1$.
Thus, $\sum_{e\in \safe\cap\delta(S)} \hat{x}_e \geq 1$.

Now suppose that $\sum_{e\in\safe\cap\delta(S)} (q+1) x^*_e +
\sum_{e\in\unsafe\cap\delta(S)} x^*_e \leq 2(q+1)$. Then, by~(P2),
the cut $\delta(S)$ satisfies the knapsack-cover inequality
w.r.t.\ the set $A = \unsafe_1$:
\[
\sum_{e\in \safe\cap\delta(S)} u_e(A,S) x^*_e +
	\sum_{e\in \unsafe_{2}\cap\delta(S)} u_e(A,S) x^*_e \geq D(A,S),
\]
where $D(A,S) = \max\{0,\, (q+1)-{\sum_{e\in\delta(S)\cap{A}} u_e}\} =
	\max\{0,\, (q+1)-|\unsafe_1\cap\delta(S)|\}$, and
$u_e(A,S) = \min\{u_e,\, D(A,S)\}$.

Moreover, since $\delta(S)$ is a small cut, we have
$\ds |\unsafe_1\cap\delta(S)| + \sum_{e\in \unsafe_{2}\cap\delta(S)} \frac{(\alpha+2)}{2} x^*_e < (q+1)$;
	thus, $(q+1) - |\unsafe_1\cap\delta(S)| \geq1$, so $D(A,S)\geq1$ and $u_e(A,S)\geq1,\forall{e}\in\delta(S)$.
We rewrite this inequality as $\ds \sum_{e\in \unsafe_{2}\cap\delta(S)} \frac{(\alpha+2)}{2} u_e(A,S) x^*_e < D(A,S)$,
which is the same as $\ds \sum_{e\in \unsafe_{2}\cap\delta(S)} u_e(A,S) x^*_e < \frac{2}{(\alpha+2)} D(A,S)$.
Thus, by the knapsack-cover inequality, we have
$\ds \sum_{e\in \safe\cap\delta(S)} u_e(A,S) x^*_e \geq \frac{\alpha}{(\alpha+2)} D(A,S)$.
By definition, $u_e(A,S) \leq D(A,S)$, hence, we have
$\ds \sum_{e\in \safe\cap\delta(S)} \frac{(\alpha+2)}{\alpha} D(A,S) x^*_e \geq D(A,S)$. 
Since $\ds \sum_{e\in \safe\cap\delta(S)} \frac{(\alpha+2)}{\alpha} x^*_e \geq 1$
and $\hat{x}_e = \min\{1,\; \frac{(\alpha+2)}{\alpha} x^*_e \}, \forall{e}\in\safe$, we have
$\sum_{e\in \safe\cap\delta(S)} \hat{x}_e \geq 1$, completing the proof.  
\end{proof}

\begin{lemma}\label{lem:oneqfgc2}
In step~3 of Algorithm~\ref{alg:oneq-FGC}, a feasible fractional
solution to the $f$-connectivity problem is given by
	$\ds x'_e = \frac{(\alpha+2)}{2} x^*_e$ for $e\in \unsafe_{2}$.
\end{lemma}

\begin{proof}
By way of contradiction, suppose that the claim does not hold.  Then
for some non-trivial cut $\delta(S)$, we have
$\ds
(q+1) |\safe_1\cap\delta(S)| +
|\unsafe_{1}\cap\delta(S)| +
	\sum_{e\in \unsafe_{2}\cap\delta(S)}  \frac{(\alpha+2)}{2} x^*_e < (q+1).
$
This implies that the cut $\delta(S)$ is a {small cut} in step~2 of
Algorithm~\ref{alg:oneq-FGC}. Hence, step~2 ensures (via $\ASC$)
that $|\safe_1\cap\delta(S)| \geq 1$ and so
$(q+1) |\safe_1\cap\delta(S)| \geq (q+1)$.
This is a contradiction.
\end{proof}

The output of Algorithm~\ref{alg:oneq-FGC} is feasible for the
$\oneqfgc$ problem by Lemmas~\ref{lem:oneqfgc1},~\ref{lem:oneqfgc2}.
The cost of the edges in $\unsafe_{1}$ is $\ds \leq \frac{(\alpha+2)}{2} \sum_{e\in
\unsafe_{1}}c_e x^*_e$ since $\ds x^*_e \geq  \frac{2}{(\alpha+2)}$ for each edge $e$ in
$\unsafe_{1}$. Additionally, the cost of the edges in $\safe_1$ is
$\leq (\alpha+2) \sum_{e\in \safe}c_e x^*_e$ by Lemma~\ref{lem:oneqfgc1}
and our definition of $\alpha$.
Lastly, the cost of the edges returned by Jain's iterative rounding
algorithm (in step~3 of Algorithm~\ref{alg:oneq-FGC}) is at most
$\ds \sum_{e\in \unsafe_{2}} (2) ((\alpha+2)/2) c_e x^*_e = (\alpha+2) \sum_{e\in
\unsafe_{2}}c_e x^*_e$, by Lemma~\ref{lem:oneqfgc2}.
Therefore, the cost of the solution returned by
Algorithm~\ref{alg:oneq-FGC} is at most $(\alpha+2) c(x^*)$.

{
In the remaining discussion of this section, let us assume that $\alpha$
is an integer and the edge costs $c$ are integers.
To ensure  $c\in\Zp^E$ (rather than  $c\in\Qp^E$), we multiply the
cost of each edge by the least common multiple of the denominators of
the edge costs (which has bit-length polynomial in the input size). This
scaling preserves the instance and the approximation ratio.

The outer loop of our algorithm runs a binary search for $\lpopt$,
but note that we are not using a ``true'' polynomial-time separation
subroutine.  Given a vector $x^*$ (a candidate solution to
\eqref{KCLP:oneqfgc}), our subroutine either finds that $x^*$
violates one of the constraints specified in
Lemma~\ref{lem:polytimeKC-oneqfgc} or else it rounds $x^*$ to an
integer solution of cost $\leq (\alpha+2) c(x^*)$, where
$c(x^*)=\sum_e{c_ex^*_e}$.
The binary search for $\lpopt$ starts
with the interval $[0,c(E)]$, where $c(E)=\sum_e{c_e}$.  
Assume that the instance has a feasible integer solution, let $\opt$
denote the cost of an optimal integer solution, and assume that
$0<\opt\leq{c(E)}$.

In an arbitrary iteration, the binary search calls the ellipsoid
algorithm with the additional constraint $\sum_e{c_ex_e}\leq{z}$,
where the current interval is $[\Lcur,\Hcur]$ and $z=\frac{\Lcur+\Hcur}{2}$.
(The binary search maintains the invariant:
$\lpopt>\Lcur$ and there exists an integer solution of cost $\leq(\alpha+2)\Hcur$.)
The ellipsoid algorithm calls our subroutine one or more times, and
either (1)~reports that the LP (with the additional constraint) is
infeasible or else (2)~it finds a vector $x^*$ with $c(x^*)\leq{z}$
and an integer solution of cost $\leq (\alpha+2) c(x^*)$.  The
binary search continues as usual, that is, in case~(1) it replaces
the current interval $[\Lcur,\Hcur]$ by the upper half-interval
$[\frac{\Lcur+\Hcur}{2},\Hcur]$, and in case~(2) it replaces the current interval
by the lower half-interval $[\Lcur,\frac{\Lcur+\Hcur}{2}]$.  
The binary search runs for $O(\log((\alpha+2)\,c(E)))$ iterations, and, when it
terminates, then the size of the current interval $[\Lfinal,\Hfinal]$ is
$\ds \Hfinal - \Lfinal ~=~ c(E) / 2^{O(\log((\alpha+2)\,c(E)))} ~<~ 1/(\alpha+2)$.
Clearly, the LP with the additional constraint $\sum_e{c_ex_e}\leq\Lfinal$
is infeasible, and the algorithm found an integer solution of cost
$\ds \leq(\alpha+2) \Hfinal$.
Hence, $\lpopt>\Lfinal$ and, moreover,
the last integer solution found by the algorithm has cost
$\ds \leq~ (\alpha+2) \Hfinal ~\leq~ (\alpha+2) (\lpopt + (\Hfinal-\Lfinal))
	~<~ (\alpha+2) (\lpopt + 1/(\alpha+2))
	~=~ (\alpha+2) \lpopt + 1 ~\leq~ (\alpha+2) \opt + 1$.
Since the edge costs are integers, we can round down the (strict) upper bound to $(\alpha+2) \opt$.
}

\oneqFGCtheorem*

}

%%%%%%%%%%
{
\section{\texorpdfstring{$O(\log{\frac{k}{u_{min}}})$}{O(log(k/u{min}))}-Approximation Algorithm for \texorpdfstring{Cap-$k$-ECSS}{Cap-k-ECSS} \label{sec:logk-approx}}

In this section, we present an $O(\log (k/u_{min}))$-approximation algorithm for
Cap-$k$-ECSS that runs in polynomial~time assuming $k/u_{min}\leq|V(G)|=n$ and $k\geq3$.
Note that when $k/u_{min}>n$, then the previously known approximation
algorithm of \cite{CCKK15} for Cap-$k$-ECSS achieves an approximation
ratio of $O(\log{n})\leq{O(\log(k/u_{min}))}$.
We assume $k\geq3$ to ensure $\lceil\log{k}\rceil\geq2$;
note that when $k<3$, then Jain's iterative rounding method \cite{Jain01,LRS-book} gives
an $O(1)$-approximation algorithm for Cap-$k$-ECSS (see section~\ref{sec:iterationT}).

Let us recall a few terms \& notation from previous sections.
For a graph $G=(V,E)$ and $S\subseteq{V}$, the cut $\delta(S)$
refers to the set of edges that have exactly one end-node in $S$;
$\delta(S)$ is called a {non-trivial} cut if $\emptyset\neq{S}\subsetneq{V}$.
(Whenever we use the term ``cut $\delta(S)$'' we mean that $S$ is a subset of $V(G)$.)
We use the term \textit{small cut} to mean a non-trivial cut
$\delta(S)$ with capacity below a specified threshold-value, say, $\lambda$.

{Let $\lpopt$ denote the optimal value of \eqref{intro-KCLP:CapkECSS},
the LP with the knapsack-cover inequalities.
Similarly to section~\ref{sec:oneqfgc}, we use binary search for
$\lpopt$ together with the ellipsoid algorithm and our polynomial-time
subroutines to find a vector $x^*$ with cost $c(x^*)=\sum_e{c_ex_e}$
such that $c(x^*) \leq \lpopt$ and $x^*$ satisfies the constraints
of \eqref{intro-KCLP:CapkECSS} specified in the proof of
Lemma~\ref{lem:polytimeKC_CapkECSS} (though $x^*$ could violate
other constraints of \eqref{intro-KCLP:CapkECSS}). Then, we will
round $x^*$ to an integer solution of cost $\leq O(\log(k/u_{min}))
c(x^*)$. Thus, we will find an integer solution of cost $\leq
O(\log(k/u_{min})) \lpopt$, even though we will not compute the
precise value of $\lpopt$.
At the end of this section, the proof of
Lemma~\ref{lem:polytimeKC_CapkECSS} discusses our overall algorithm in more detail.
}

\noindent
\textbf{Assumption}:
In what follows, assume that the vector $x^*$ satisfies all the constraints of \eqref{intro-KCLP:CapkECSS}.
In section~\ref{sec:solvingLP} below, we explain that we can easily remove this assumption.

%-----
\begin{figure}[htb]
{
\begin{tikzpicture}[scale=0.90]

    % Define page width for scaling
    \def\PageWidth{14}
    
    % Draw main line
    \draw[thick] (0,0) -- (\PageWidth,0);

    % Define coordinates
    \coordinate (A) at (0,0);
    \coordinate (B) at (3,0);
    \coordinate (C) at (6,0);
    \coordinate (D) at (12,0);
    \coordinate (E) at (\PageWidth,0);
    \coordinate (F) at (13.5,0); % Position for E_{T-1}

    % Draw expanded ovals around sections
    \draw[blue, thick, rounded corners=18pt] (-0.5,-0.8) rectangle (3.5,0.8);  % Oval for E1
    \draw[blue, thick, rounded corners=18pt] (-0.5,-1) rectangle (6.5,1.2);  % Oval for E2
    \draw[blue, thick, rounded corners=18pt] (-0.5,-1.5) rectangle (\PageWidth+0.5,2); % Oval for ET
    \draw[blue, thick, rounded corners=18pt] (-0.5,-1.2) rectangle (12.7,1.5); % Oval for E_{T-1}

    % Add text labels
    \node at (A) [below] {$0$};
    \node at (B) [below] {$2^1$};
    \node at (C) [below] {$2^2$};
    \node at (D) [below] {$2^{T-1}$};
    \node at (E) [below] {$2^{T}$};
    \node at (9,-0.3) {...};

    % Add E labels in ovals
    \node at (1.5,0.5) {\textcolor{blue}{$E_1$}};
    \node at (4.5,0.9) {\textcolor{blue}{$E_2$}};
    \node at (12.5,2.3) {\textcolor{blue}{$E_T$}};
    \node at (11.5,1.2) {\textcolor{blue}{$E_{T-1}$}};

    % Curly braces for intervals
    \draw[thick,decorate,decoration={brace,amplitude=8pt,mirror}] (3.2,-1.6) -- (6.2,-1.6) node[midway,yshift=-0.6cm] {$E_2 - E_1$};
    \draw[thick,decorate,decoration={brace,amplitude=8pt,mirror}] (10.5,-1.6) -- (12.5,-1.6) node[midway,yshift=-0.6cm,xshift=-0.2cm] {$E_{T-1} - E_{T-2}$};
    \draw[thick,decorate,decoration={brace,amplitude=8pt,mirror}] (12.5,-1.6) -- (\PageWidth+0.3,-1.6) node[midway,yshift=-0.6cm,xshift=0.2cm] {$E_T - E_{T-1}$};

    % Dotted (not Dashed) line for omitted values
    \draw[thick,blue,dotted] (7,0.5) -- (11,0.5);

\end{tikzpicture}
}
\caption{\label{fig:E-buckets} Illustration of buckets. Note that $T=\lceil\log{k}\rceil$.}
\end{figure}
%-----

Throughout the execution of the rounding algorithm, we will maintain a set of
edges $E_{cur}$ acting as our current solution.
We begin with $E_{cur} = \{e\in E:x^*_e \geq 1/2\}$.
Define $T = \lceil\log{k}\rceil$ and
for $j=1,2,\ldots,T$,
define $E_j = \{e\in E: x^*_e < 1/2 \text{ and } u_e \leq 2^j\}$;
thus, the edge-sets
$E_T-E_{T-1}, E_{T-1}-E_{T-2}, \ldots, E_3-E_2, E_2-E_1, E_1$
form a partition of the edges in $E - E_{cur}$ into $T$ buckets
based on the capacities;
let us call the set $E_{T-i+1} - E_{T-i}$ the $i$-th bucket
(and $E_1$ is the $T$-th bucket).
See Figure~\ref{fig:E-buckets} for an illustration.

Our algorithm will have $T$ iterations and each iteration (except
for the first and the last) will have two phases.
During phase~1 of iteration~$i$,
we will round some of the edges in the $i$-th bucket,
i.e., some of the edges in the set $E_{T-i+1} - E_{T-i}$.
Note that an edge $e$ in the $i$-th bucket has capacity $2^{T-i}<u_e\leq2^{T-i+1}$.
Informally speaking, in phase~1, we want to augment cuts of very small capacity with edges of capacity $\approx2^{T-i}$,
and, in general, we need $\Theta(k/2^{T-i})$ rounds of augmentation to achieve capacity $k$;
thus, phase~1 has $\Theta(k/2^{T-i})$ sub-iterations.
During phase~2 of iteration~$i$,
we will round some of the edges in $E-(E_{T-i}\cup{E_{cur}})$.

Next, we present pseudo-code for the rounding algorithm, followed
by explanation and analysis of the main steps.

{
\begin{algorithm}[H]
    \caption{$O(\log(k/u_{min}))$-approximate solution to Cap-$k$-ECSS}\label{alg:capkECSS}

    \textbf{Require:} Graph $G = (V,E)$ with capacities $\{u_e\}_{e \in E}$ and costs $\{c_e\}_{e\in E}$.  A vector $x^*$ satisfying some constraints of \eqref{intro-KCLP:CapkECSS}, set $E_{cur} = \{e\in E:x^*_e \geq 1/2\}$, and sets $E_j \subseteq E, j = 1,\dots,\lceil \log k \rceil$ as defined above.
\smallskip
    
    \begin{enumerate}
	\setcounter{enumi}{-1}

        \item  \textbf{In any iteration, at the start, if $E_{cur}$ is feasible,
			then stop and return $E_{cur}$.}

        \item  Iteration 1: Apply two iterations of the following steps~(a), (b).
        \begin{enumerate}[(a)]
        \item Let $\C = \{\emptyset\neq S \subsetneq V \; : \; \sum_{e \in E_{cur} \cap \delta(S)}u_e + \sum_{e \in E_{T-1} \cap \delta(S)} 2 u_e x^*_e < k\}$.  
        
        \item  Apply the approximation algorithm for $\ASC$ to select edges from $E - E_{T-1} - E_{cur}$ to cover the cuts in $\C$.  Add the selected edges to $E_{cur}$.

%       \item Repeat (a), (b) once.
    \end{enumerate}

    \item For $i = 2,\dots,T-1$, Iteration $i$:
    \begin{enumerate}[(a)]
        \item For $\ell = 1,\dots,\lfloor (k-2^{T-i+1})/2^{T-i} \rfloor$:
        \begin{enumerate}[(i)]
            \item Let $\C = \{\emptyset\neq S \subsetneq V \; : \; \sum_{e \in E_{cur} \cap \delta(S)}u_e + \sum_{e \in E_{T-i} \cap \delta(S)} 2 u_e x^*_e < \ell\; 2^{T-i}\}$.  

            \item Apply the approximation algorithm for $\ASC$ to select edges from $E _{T-i+1}- E_{T-i} - E_{cur}$ to cover the cuts in $\C$.  Add the selected edges to $E_{cur}$.
        \end{enumerate}

        \item Apply three iterations of the following steps~(c), (d).

        \item Let $\C = \{\emptyset\neq S \subsetneq V \; : \; \sum_{e \in E_{cur} \cap \delta(S)}u_e + \sum_{e \in E_{T-i} \cap \delta(S)} 2 u_e x^*_e < k\}$.  
            
        \item Apply the approximation algorithm for $\ASC$ to selected edges from $E - E_{T-i} - E_{cur}$ to cover the cuts in $\C$.  Add the selected edges to $E_{cur}$.

%       \item Repeat (b), (c) \textbf{two} additional times.
    \end{enumerate}

    \item Iteration $T$:
    \begin{enumerate}[(a)]
        \item At this point, we have that $\sum_{e \in E_{cur} \cap \delta(S)}u_e + \sum_{e \in E_1 \cap \delta(S)}2 u_e x^*_e \geq k$ for all $S \subsetneq V, S \neq \emptyset$.  Apply Jain's iterative rounding method to round the ($x$~variables of the) edges in $E_1$ to an integer solution $E_1^*$, such that $E_{cur} \cup E_1^*$ is a feasible solution to Cap-$k$-ECSS.
    \end{enumerate}

    \item Return $E_{cur} \cup E_1^*$.
    \end{enumerate}
\end{algorithm}
}

%-----
{
For every non-trivial cut $\delta(S)$, we will maintain the following invariants for all iterations $i = 2,\dots,(T-1)$ (i.e., except the first and the last iteration):

\begin{enumerate}[(1)]
    \item At the beginning of iteration $i$, $E_{cur} \cap E_{T-i+1} = \emptyset$ and
	$$\sum_{e\in E_{cur} \cap \delta(S)}u_e + \sum_{e\in E_{T-i + 1} \cap \delta(S)} {2}\ u_ex^*_e \geq k.$$
We note that iteration~$1$ ensures that this invariant holds at the start of iteration $2$.

    \item At the end of phase~1 of iteration $i$,
    \[
    \sum_{e\in E_{cur} \cap \delta(S)}u_e + \sum_{e\in E_{T-i} \cap \delta(S)}{2}\  u_ex^*_e \geq k-2^{T-i+1} - 2^{T-i}.
    \]

    \item At the end of iteration $i$, which is also the end of phase~2
    of iteration $i$, $E_{cur} \cap E_{T-i} = \emptyset$, and
    \[\sum_{e\in E_{cur} \cap \delta(S)}u_e + \sum_{e\in E_{T-i} \cap \delta(S)} {2}\  u_ex^*_e \geq k.\]

\item[]
Observe that invariant~(3) for iteration~$i$ is the same as
invariant~(1) for iteration~$i+1$.
\end{enumerate}

\begin{remark}
Our analysis of Steps~1\,(b) and~2\,(d) of Algorithm~\ref{alg:capkECSS}
is similar to our analysis of Step~2\,(a) of Algorithm~\ref{alg:oneq-FGC},
and uses the knapsack-cover inequalities. Our analysis applies for
each of the (two or three) sub-iterations with the \textit{updated} $E_{cur}$ and $A=E_{cur}$,
where $A$ denotes the picked edge-set for the knapsack-cover inequalities.
\end{remark}
}
%-----

\subsection{Iteration 1}

In this iteration, we consider the family of small~cuts $\delta(S)$ where
\[
\sum_{e\in E_{cur} \cap \delta(S)}u_e+ \sum_{e\in E_{T-1} \cap \delta(S)}{2}\  u_ex^*_e < k \tag{definition of small cuts}
\]
We will cover these cuts using edges in $E-E_{T-1} - E_{cur}$.
Consider any one of these small cuts $\delta(S)$. 
Since $\delta(S)$ is a small cut, we have
$\sum_{e\in E_{T-1} \cap \delta(S)}{2}\ u_ex^*_e <
	k - (\sum_{e\in E_{cur}\cap\delta(S)} u_e) = k - u(E_{cur}\cap\delta(S))$,
hence, we have $\sum_{e\in E_{T-1} \cap \delta(S)}u_ex^*_e <
	(k- u(E_{cur}\cap\delta(S)))/{2}$.
Consider the knapsack-cover inequality for one of these small~cuts
$\delta(S)$ and the set $A=E_{cur}$,
	$\sum_{e\in{E\cap\delta(S)-A}} u_e(A,S)\,x_e\geq{D(A,S)}$, where
	$D(A,S)=(k - u(E_{cur}\cap\delta(S)))$ and $u_e(A,S)=\min\{u_e,D(A,S)\}$.
By the above inequality and the knapsack-cover inequality,
each of these small~cuts $\delta(S)$ satisfies the inequality
$\sum_{e\in (E-E_{T-1} - E_{cur}) \cap \delta(S)} \min\{u_e,D(A,S)\} x^*_e > D(A,S)/{2}$,
which implies the inequality
	$\sum_{e\in (E-E_{T-1} - E_{cur}) \cap \delta(S)} D(A,S) x^*_e > D(A,S)/{2}$.
Thus ${2}\,x^*_{E-E_{T-1} - E_{cur}}$ is feasible for the
$\ASC$ problem implying that we incur a cost of at most
$\pliableapx \cdot {2} \cdot c(x^*_{E_T-E_{T-1}})$ here.
We run the $\pliableapx$-approximation algorithm for $\ASC$,
see Proposition~\ref{prop:approxASC},
and use the edge-set returned by that algorithm to augment $E_{cur}$.
We repeat one more time, i.e., we again consider all cuts $\delta(S)$ where
$\sum_{e\in E_{cur} \cap \delta(S)}u_e+ \sum_{e\in E_{T-1} \cap
\delta(S)}{2} u_ex^*_e < k$ and cover these cuts using edges
in $E-E_{T-1} - E_{cur}$, incurring a further cost of $\pliableapx \cdot
{2} \cdot c(x^*_{E-E_{T-1}})$.
Now, we will have necessarily satisfied invariant~(3) at the end
of this iteration.  To see this, observe that if some cut violated
this invariant, then this cut participated as a small~cut in both
instances of $\ASC$ considered in this step.  This means we would
have added at least two edges that cover this cut, each of capacity
at least $k/2$, ensuring that invariant~(3) holds.

\subsection{Iteration \texorpdfstring{$i$}{i} Phase 1 (Step 2 (a) in Algorithm \ref{alg:capkECSS})}

{
We are starting with invariant~(1) at the beginning of this iteration
(as this corresponds to the invariant~(3) that holds at the end of
the previous iteration). Hence we have $E_{cur} \cap E_{T-i+1} =
\emptyset$ and
$\sum_{e\in E_{cur}\cap\delta(S)}u_e + \sum_{e\in E_{T-i + 1}\cap\delta(S)}
{2} u_ex^*_e \geq k$. We will run multiple sub-iterations
within this phase. The first sub-iteration is described below.
        
Consider the family of small~cuts $\delta(S)$ where $\sum_{e\in
E_{cur} \cap \delta(S)}u_e + \sum_{e\in E_{T-i} \cap \delta(S)}
{2} u_ex^*_e < 2^{T-i}$. We will cover these cuts using edges 
in $E_{T-i+1}-E_{T-i}-E_{cur}$.  For these small~cuts $\delta(S)$, we have
$$\sum_{e\in (E_{T-i+1}-E_{T-i}-E_{cur}) \cap \delta(S)}{2} u_ex^*_e \geq k-2^{T-i}$$
\big(this inequality is obtained by subtracting the inequality defining
the small~cuts from the inequality of invariant~(1);
let $E_{cur}^{(0)}$ denote $E_{cur}$ at the start of iteration~$i$;
the inequality resulting from the subtraction is
$\sum_{e\in (E_{cur}-E_{cur}^{(0)}) \cap \delta(S)} ({2} u_ex^*_e - u_e) \;+\;
	\sum_{e\in (E_{T-i+1}-E_{T-i}-E_{cur}) \cap \delta(S)}{2} u_ex^*_e \geq k-2^{T-i}$,
and the first term is non-positive since $x^*_e\leq1/2,\forall{e}\in{E_{cur}-E_{cur}^{(0)}}$\big).
Observe that
$\sum_{e\in (E_{T-i+1}-E_{T-i}-E_{cur}) \cap \delta(S)}{2} x^*_e $ $\geq$
$(k-2^{T-i})/2^{T-i+1}$, because $u_e \leq 2^{T-i+1}$ for all edges in $E_{T-i+1}$.
Thus, ${2} x^*_{E_{T-i+1}-E_{T-i}-E_{cur}} \cdot 2^{T-i+1}/(k-2^{T-i})$
is feasible for the $\ASC$ problem.
We run the $\pliableapx$-approximation algorithm for $\ASC$,
see Proposition~\ref{prop:approxASC},
and use the edge-set returned by that algorithm to augment $E_{cur}$.
After this, there are no non-trivial cuts $\delta(S)$ with
$\sum_{e\in E_{cur} \cap \delta(S)}u_e + \sum_{e\in E_{T-i} \cap \delta(S)}
{2} u_ex^*_e < 2^{T-i}$ since we would have covered any such
cut by an edge from $E_{T-i+1}-E_{T-i}-E_{cur}$, and all these edges
have capacity at least $2^{T-i}$. Next, we shift the threshold in
the definition for small~cuts to $2\cdot 2^{T-i}$, and then to
$3\cdot 2^{T-i}$, \dots, all the way until $\hat{\ell}\cdot 2^{T-i}$
where $\hat{\ell} = \lfloor (k-2^{T-i+1})/2^{T-i} \rfloor$. This
would imply that  $\hat{\ell}\cdot2^{T-i}\geq k-2^{T-i+1} - 2^{T-i}$.
We describe these sub-iterations in more detail now.

{
For $\ell = 1,2,\ldots, \hat{\ell}$, consider the family of small~cuts $\delta(S)$ where
\[
\sum_{e\in E_{cur} \cap \delta(S)}u_e + \sum_{e\in E_{T-i} \cap \delta(S)} {2} u_ex^*_e < \ell\cdot 2^{T-i}. \tag{definition of small cuts}
\]
Since invariant~(1) is true (and we have only increased the LHS of
invariant~(1) during the phase), we have
{$$\ds
\sum_{e\in (E_{T-i+1}-E_{T-i}-E_{cur}) \cap \delta(S)}{2}  u_ex^*_e \geq k-\ell\cdot 2^{T-i}.
$$}
Since $u_e \leq 2^{T-i+1}$ for all edges in $E_{T-i+1}$, we have
$\ds
\sum_{e\in (E_{T-i+1}-E_{T-i}-E_{cur}) \cap \delta(S)}{2}  x^*_e \geq (k-\ell\cdot 2^{T-i})/2^{T-i+1}.$
Thus, ${2} x^*_{E_{T-i+1}-E_{T-i}-E_{cur}} \cdot 2^{T-i+1}/(k-\ell\cdot 2^{T-i})$
is feasible for our instance of $\ASC$.
We run the $\pliableapx$-approximation algorithm for $\ASC$,
see Proposition~\ref{prop:approxASC},
and use the edge-set returned by that algorithm to augment $E_{cur}$.
Then, we move on to the next sub-iteration. At the end of the last sub-iteration (with $\ell=\hat{\ell}$), 
we have
\[
    \sum_{e\in E_{cur} \cap \delta(S)}u_e + \sum_{e\in E_{T-i} \cap \delta(S)}{2} u_ex^*_e \geq (\hat{\ell})2^{T-i} \geq k-2^{T-i+1}-2^{T-i},
\]
and so invariant~(2) is maintained. Let us analyze the cost we incurred in this phase.
}

The cost we incur is at most
$\pliableapx \cdot {2} \cdot  c(x^*_{E_{T-i+1}-E_{T-i}}) \cdot 2^{T-i+1} \cdot
\big(\frac{1}{k-2^{T-i}} + \frac{1}{k-2\cdot2^{T-i}} + \cdots+\frac{1}{k-\hat{\ell}2^{T-i}}\big)$.
We bound this last sum as follows. Note that $\hat{\ell} 2^{T-i} \leq k - 2^{T-i+1}$ and so
$k-\hat{\ell}2^{T-i} \geq 2^{T-i+1}$.
\begin{align*}
    &\frac{1}{k-2^{T-i}} + \frac{1}{k-2\cdot2^{T-i}} +\cdots+ \frac{1}{k-\hat{\ell}2^{T-i}} \\
    &= \frac{1}{k-\hat{\ell}2^{T-i}} + \frac{1}{k-\hat{\ell}2^{T-i} + 2^{T-i}}+ \frac{1}{k-\hat{\ell}2^{T-i} + 2\cdot 2^{T-i}} +\cdots+ \frac{1}{k-\hat{\ell}2^{T-i} + (\hat{\ell}-1)\cdot 2^{T-i}}\\
    &\leq \frac{1}{2^{T-i+1}} + \sum_{\ell=1}^{\hat{\ell}-1} \frac{1}{k-\hat{\ell}2^{T-i} + 2^{T-i}\ell}\\
    &\leq \frac{1}{2^{T-i+1}} + \int_0^{\hat{\ell}-1} \frac{1}{k-\hat{\ell}2^{T-i} + 2^{T-i}\ell} d\ell \\
    & = \frac{1}{2^{T-i+1}}+ \frac{1}{2^{T-i}}\left(\log(k-2^{T-i}) - \log(k-\hat{\ell}2^{T-i})\right)\\
    &\leq \frac{1}{2^{T-i+1}}+ \frac{1}{2^{T-i}}\left(\log(k) - \log(2^{T-i+1})\right)
        \tag{using the inequality $k-\hat{\ell}2^{T-i}\geq2^{T-i+1}$} \\
    & {= O\left(\frac{\log(k/2^{T-i+1})}{2^{T-i}}\right)} \\
\end{align*}
Thus, the cost incurred in this phase is
$\leq \pliableapx \cdot {2} \cdot (2^{T-i+1}/2^{T-i}) \cdot {O(\log(k/2^{T-i+1}))} \cdot
	c(x^*_{E_{T-i+1}-E_{T-i}})$.
}

\subsection{Iteration \texorpdfstring{$i$}{i} Phase 2 (Step 2 (b)-(d) in Algorithm \ref{alg:capkECSS})}

We are beginning with invariant~(2), which is valid at the end of
phase~1, and thus for all non-trivial cuts $\delta(S)$, we have
$\ds
    \sum_{e\in E_{cur} \cap \delta(S)}u_e + \sum_{e\in E_{T-i} \cap \delta(S)}{2} u_ex^*_e \geq k-2^{T-i+1} - 2^{T-i}.$
We will add more edges from $E-E_{T-i}-E_{cur}$ to these cuts, if
needed, to increase the capacity to $k$. Note that all edges in
$E-E_{T-i}-E_{cur}$ have capacity at least $2^{T-i}$ and so at most
three more edges need to be added. To do so, we employ the method
we used in iteration~1.

Consider the family of small~cuts $\delta(S)$ where
$\ds \sum_{e\in E_{cur} \cap \delta(S)} u_e + \sum_{e\in E_{T-i} \cap \delta(S)}{2} u_e x^*_e <k.$
Then, by the knapsack-cover inequalities,
${2}\,x^*_{E-E_{T-i}-E_{cur}}$ is feasible for the $\ASC$
instance. Indeed for each of these small~cuts $\delta(S)$, we have
$\sum_{e\in E_{T-i} \cap \delta(S)}u_ex^*_e <
(k - \sum_{e\in E_{cur}\cap\delta(S)}u_e)/{2} = (k - u(E_{cur}\cap\delta(S)))/{2}$.
The knapsack-cover inequality then implies that
$\sum_{e\in (E-E_{T-i} - E_{cur}) \cap \delta(S)} D(A,S) x^*_e > D(A,S)/{2}$,
where $A=E_{cur}$ and $D(A,S) = k - u(E_{cur}\cap\delta(S))$.
We run the $\pliableapx$-approximation algorithm for $\ASC$,
see Proposition~\ref{prop:approxASC};
and use the edge-set returned by that algorithm to augment $E_{cur}$.
This incurs a cost of at most $\pliableapx\cdot{2}\cdot c(x^*)$.
We repeat three times, adding the approximate solution of the $\ASC$
instance to $E_{cur}$ and incur a cost of at most $3\cdot\pliableapx\cdot{2}\cdot c(x^*)$.
At the end of this phase, for every non-trivial cut $\delta(S)$, we have
$\ds \sum_{e\in E_{cur} \cap \delta(S)}u_e + \sum_{e\in E_{T-i} \cap \delta(S)}{2} u_e x^*_e \geq k.$
This is precisely invariant~(3) and we have completed this phase.

\subsection{Iteration \texorpdfstring{$T$}{T}} \label{sec:iterationT}

At the beginning of the last iteration, by invariant~(1), we have for any non-trivial cut $\delta(S)$:
$$
 \sum_{e\in E_{cur} \cap \delta(S)}u_e + \sum_{e\in E_{1} \cap \delta(S)}{2} u_e x^*_e \geq k.$$
Now, we apply Jain's iterative rounding method \cite{Jain01,LRS-book} to round the edges
in $E_1$, incurring a cost of at most
$(2\;\max\{u_e: e\in E_1\})\cdot c(x^*_{E_1}) = 4c(x^*_{E_1})$.
\big(Jain's analysis extends to the setting of positive integer capacities,
via the variable substitution $x'_e = u_e x_e$, with
approximation ratio $2$ times the maximum edge capacity.\big)

\subsection{Total Cost}
{
We calculate the total cost incurred separately for each iteration.
In iteration 1, we incur a cost of $O(1)\cdot c(x^*)$;
this includes the term $\sum_{\{e\in{E}\,:\,x^*_e\geq1/2\}} 2 c_e x^*_e$
(due to the initial $E_{cur}$).
In phase~1
of iterations $i = 2,\ldots, T-1$, we incur a total cost of
\[
O(1)\cdot \sum_{i=2}^{T-1}\log(k/2^{T-i+1}) c(x^*_{E_{T-i+1} - E_{T-i}}) 
    ~~\leq~~ O(\log k)c(x^*).
\]
In phase~2 of iterations $i=2,\ldots, T-1$, we incur a cost of
$\sum_{i=2}^{T-1} O(1)\cdot c(x^*) \leq O(T) c(x^*) \leq O(\log k)
c(x^*)$. Finally the cost incurred in iteration $T$ is $O(1)\cdot
c(x^*)$. Thus the total cost incurred is $O(\log k) c(x^*)$.

Observe that if the minimum capacity $u_{min}$ over all edges in
$E$ is greater than 2, then the algorithm stops at an earlier iteration.
Let $\ifinal$ denote the value of the index~$i$ such that
$2^{T-\ifinal} < u_{min} \leq 2^{T-\ifinal+1}$.
Since $T = \lceil \log k \rceil$, we have $\ifinal = O(\log (k/u_{min}))$.
Observe that $E_{T-\ifinal}$ is empty.
At the end of iteration $\ifinal$, by invariant~(3),
for every non-trivial cut $\delta(S)$, we have
    \[\sum_{e\in E_{cur} \cap \delta(S)}u_e + \sum_{e\in E_{T-\ifinal} \cap \delta(S)} {2}\  u_ex^*_e \:=\: u(E_{cur} \cap \delta(S)) \geq k.\]
Hence, $E_{cur}$ is feasible for the instance, and 
(at the start of the next iteration) the algorithm stops and outputs $E_{cur}$.

In such a scenario, the total cost
incurred in phase~1 of iterations $i = 2,\ldots, \ifinal$, is
\begin{align*}
    &O(1)\cdot \sum_{i=2}^{\ifinal}\log(k/2^{T-i+1}) c(x^*_{E_{T-i+1} - E_{T-i}}) \\
    &\leq O(1)\cdot \sum_{i=2}^{\ifinal} \log(k/u_{min}) c(x^*_{E_{T-i+1} - E_{T-i}}) \tag{since $u_{min} \leq 2^{T-\ifinal+1}$}\\
    &\leq O(\log(k/u_{min}))c(x^*).
\end{align*}
Similarly the total cost incurred in phase~2 of iterations $i =
2,\dots,\ifinal$ is $\sum_{i=2}^{\ifinal} O(1)\cdot c(x^*) \leq
O(\ifinal) c(x^*) \leq O(\log (k/u_{min})) c(x^*)$. Thus the overall
cost is $O(\log(k/u_{min})) c(x^*)$.
}

\subsection{Solving the LP Relaxation \label{sec:solvingLP}} 

Clearly, our rounding algorithm runs in polynomial~time, provided 
an optimal (and feasible) solution $x^*$ to \eqref{intro-KCLP:CapkECSS} is given. 
As in section~\ref{sec:oneqfgc}, we would like solve \eqref{intro-KCLP:CapkECSS}
using the ellipsoid method, but, unfortunately, we do not know of any 
polynomial-time separation oracle for the entire set of knapsack-cover inequalities.
Instead, we will {iteratively} (in polynomial time) identify a subset 
of edges $A$ and a collection of sets $\C$
such that, as long as the knapsack-cover inequalities hold for $A$ and all $S\in\C$,
we will be able to execute our rounding algorithm.

\begin{lemma}\label{lem:polytimeKC_CapkECSS}
There is a polynomial-time algorithm that,
given a vector $x^*$ (that is a candidate solution of \eqref{intro-KCLP:CapkECSS})
and a value $z$,
either finds a violated constraint of the LP
or else verifies that
$c(x^*)\leq{z}$ and, moreover, 
for every iteration $i$, $i=1,\dots,(T-1)$,
$x^*$ satisfies the property that $2x^*_{E-E_{T-i} - E_{cur}}$ is feasible
for the LP relaxations of the $\ASC$ instances created in steps~1(b) and~2(d) of Algorithm~\ref{alg:capkECSS}.
\end{lemma}
\begin{proof}
Given a candidate vector $x^*$ and a candidate objective value $z$, we first check that $\sum_{e \in E} c_e x_e^* \le z$ (see the discussion on binary search at the end of section~\ref{sec:oneqfgc}).
If not, we return this as a violated constraint. Otherwise, let $\hG=(\hV,\hE)$ be the capacitated graph where $\hV=V,\hE=E,$ and each edge $e \in \hE$ is assigned a capacity of $u_ex_e^*$. We can now check that the capacity of a minimum~cut in $\hG$ is at least $k$ using a polynomial-time global minimum-cut algorithm \cite{Schrijver-book}.  If not, we return a global minimum cut in $\hG$ as a violated constraint. 

By Karger's result \cite{Karger93}, we know that there are at most
$O(n^4)$ cuts of capacity at most $2k$ (i.e., at most twice the
capacity of a minimum~cut), and, moreover, we can enumerate all
such cuts of $\hG$ in polynomial time \cite{NNI97}.
By iterating over each of the $O(n^4)$ cuts, we can then verify in
polynomial~time that the knapsack-cover inequalities are satisfied
w.r.t.\ the set $A = E_{cur}$ in each of the steps~1(b) and~2(d)
for cuts whose capacity is at most $2k$. If not, we have found a
violated constraint. It remains then to handle the case when we are
at step~1(b) or~2(d), and we have a small~cut $\delta(S)$ such
that $ \sum_{e \in \delta(S)} u_e x^*_e > 2k$;
let us call $ \sum_{e \in \delta(S)} u_e x^*_e $ the fractional capacity of the cut.

In this case, we note that in step~1(b), by the definition of
small~cuts, we have
{$$\sum_{e\in E_{cur} \cap \delta(S)}u_e \;+\;
		\sum_{e\in E_{T-1} \cap \delta(S)} {2} u_ex^*_e < k.$$}
But then since the fractional capacity of this cut is at least $2k$, it
follows that {$$\sum_{e\in (E-E_{T-1} - E_{cur}) \cap \delta(S)} u_e x^*_e > k.$$}
Since $u_e \le k$ for every edge $e \in E$, it follows that $\sum_{e\in
(E-E_{T-1} - E_{cur}) \cap \delta(S)} x^*_e > 1$. Thus
${2}\,x^*_{E-E_{T-1} - E_{cur}}$ is feasible on this cut for
the $\ASC$ instance. A similar argument can be used to show that
in step 2(d), if we have a small~cut $\delta(S)$ with $ \sum_{e \in
\delta(S)} u_e x^*_e > 2k,$ then ${2}\,x^*_{E-E_{T-i}-E_{cur}}$
is feasible for the $\ASC$ instance.
Specifically, by the definition of small~cuts, we have
$\sum_{e \in E_{cur} \cap \delta(S)}u_e + \sum_{e \in E_{T-i} \cap \delta(S)} 2 u_e x^*_e < k$.
As the fractional capacity of $\delta(S)$ is at least $2k$, it follows
that $\sum_{e\in (E-E_{T-i} - E_{cur}) \cap \delta(S)} u_ex^*_e > k$, and
since $u_e \leq k$ for every edge $e \in E$,
$\sum_{e\in (E-E_{T-i} - E_{cur}) \cap \delta(S)} x^*_e > 1$.  
Thus, for this cut $\delta(S)$, $2x^*_{E - E_{T-i} - E_{cur}}$ is feasible for the $\ASC$ instance.

Finally, if at any step of the rounding algorithm, we identify a
violated constraint, then we \textit{restart} the rounding algorithm
from the {very beginning}. It is worth highlighting that the
verification of knapsack-cover inequalities identified
in steps~1(b) and~2(d) of the algorithm, is always done with respect
to the solution $x^*$ given by the Ellipsoid algorithm (without any
modification). As the rounding progresses, the only thing that
changes is the definition of the set $A = E_{cur}$ with respect to
which we verify the knapsack-cover inequalities. So whenever a
violated constraint is identified, it contributes to the iteration
count of the ellipsoid algorithm. Since the ellipsoid algorithm
terminates after {$n^{O(1)}$} iterations of feasibility verification
\cite{ellipsoid-book}, it must be the case that after at most
{$n^{O(1)}$} re-starts of the rounding process, we arrive at a
solution $x^*$  to \eqref{intro-KCLP:CapkECSS} of value at most
$\lpopt$ such that the solution satisfies the property that
$2x^*_{E-E_{T-i} - E_{cur}}$ is feasible for the $\ASC$ instances
created in steps~1(b) and~2(d).
\end{proof}

\begin{remark}
We mention that the analysis of phase~1 of iteration~$i$ $(i=2,\dots,(T-1))$ (i.e., step~2(a) of iteration~$i$)
does \textit{not} use the knapsack-cover inequalities, hence, Lemma~\ref{lem:polytimeKC_CapkECSS} does not address step~2(a).
Recall that phase~1 of iteration~$i$ $(i=2,\dots,(T-1))$ relies on invariant~(1), which is
established by phase~2 of the previous iteration.
\end{remark}

\CapkECSStheorem*

}
%%%%%%%%%%
{
\section{\texorpdfstring{$O(1)$}{O(1)}-Approximate Reductions between \texorpdfstring{$\twoqfgc$}{(2,q)-FGC} and \texorpdfstring{$\twoASC$}{2-Cover Small Cuts} \label{sec:twoqfgc}}

We provide reductions between the problems $\twoqfgc$ and $\twoASC$ in both directions.
Each of these reductions preserves the approximation ratio up to a constant factor.

Recall that in the $\twoASC$ problem,
we are given a graph $\tG=(\tV,\tE)$ with edge capacities $u$, a number $\capbound$,
as well as a set of links $L \subseteq \binom{\tV}{2}$ with link costs $c$;
the goal is to find a cheapest set of links $\J\subseteq{L}$
that two-covers the family of small cuts,
namely, $\{S\subsetneq{\tV}, S\not=\emptyset \mid u(\delta(S)) < \capbound\}$.

\subsection{Reducing \texorpdfstring{$\twoqfgc$}{(2,q)-FGC} to \texorpdfstring{$\twoASC$}{2-Cover Small Cuts} with approximation ratio \texorpdfstring{$O(1)$}{O(1)}}
{
Suppose we have an LP~relative $\rho$-approximation algorithm for $\twoASC$; assume $\rho\geq1$.
{
(In other words, suppose we have access to a polynomial-time algorithm that rounds
a given fractional feasible solution of the LP relaxation of $\twoASC$
to a feasible (integral) solution of cost $\leq \rho\,\lpopt$, 
where $\lpopt$ denotes the optimal value of the LP~relaxation.)
}

Let $G=(V,E = \safe \cup \unsafe)$ be an instance of $\twoqfgc$
with edge costs  $c \in \Qp^E$.  We use the following LP~relaxation.
Informally speaking, the first type of constraints (see~(1) below) correspond to
a $\oneqplusfgc$ problem,
and the second type of constraints (see~(2) below) state that for each safe edge $f$,
the edges in $E-f$ should satisfy the requirements of $\oneqfgc$
(i.e., each non-trivial cut has one safe edge or $q+1$ edges).
One can verify that the incidence vector of any feasible solution of $\twoqfgc$
satisfies all the constraints of this LP.
\begin{align*}\label{LP:twoqfgc}
    \min\;\;&\sum_{e\in E}c_e x_e \tag{LP:$(2,q)$-FGC}\\
    \text{s.t.}\;\;& \sum_{e\in \safe \cap \delta(S)} (q+2)x_e + \sum_{e\in \unsafe \cap\delta(S)}x_e \geq q+2 && \forall\;\; \emptyset\subsetneq S\subsetneq V\tag{1} \\
    & \sum_{e\in (\safe - f)\cap \delta(S)} (q+1)x_e + \sum_{e\in \unsafe \cap\delta(S)}x_e \geq q+1 && \forall\;\; \emptyset\subsetneq S\subsetneq V,\;\;\forall\;\; f\in\safe \tag{2}\\
    &0\leq x\leq 1
\end{align*}
Each of the inequalities~(1) and each of the inequalities~(2) (for
each $f\in\safe$, $\forall S\subsetneq{V},S\not=\emptyset$) can be
strengthened using the knapsack-cover inequalities.
Let us denote the stronger LP (with the knapsack-cover inequalities) by ({KCLP:$\,(2,q)$-FGC}).

{
We follow the method of section~\ref{sec:oneqfgc} for rounding this strengthened LP.
Let $\lpopt$ denote the optimal value of ({KCLP:$\,(2,q)$-FGC}), 
the LP with the knapsack-cover inequalities.  Using binary search
for $\lpopt$ together with the ellipsoid algorithm and our
polynomial-time subroutines, we will find a vector $x^*$ with
cost $c(x^*)=\sum_e{c_ex_e}$ such that $c(x^*) \leq
\lpopt$ and $x^*$ satisfies the constraints of ({KCLP:$\,(2,q)$-FGC}) 
specified in the next lemma (though $x^*$ could violate other
constraints of ({KCLP:$\,(2,q)$-FGC})).
As in section~\ref{sec:oneqfgc}, we will design a polynomial-time
algorithm for rounding the strengthened LP by picking some sets of
unsafe edges, and a polynomial-time computable collection of sets 
$\C$.  We will show that our rounding algorithm succeeds as long 
as the knapsack-cover inequalities hold for each set $S\in\C$ and
an appropriate set of unsafe edges $A$. 
}

{
\begin{lemma}\label{lem:polytimeKC-twoqfgc}
There is a polynomial-time algorithm that given a vector $x^*$ 
(that is a candidate solution of ({KCLP:$\,(2,q)$-FGC})) and
a value $z$
either finds a violated constraint of the LP
or else verifies that $c(x^*)\leq{z}$ and, moreover, $x^*$ satisfies the following properties:
\begin{itemize}
\item[(P3)]
$\sum_{e\in\safe\cap\delta(S)} (q+2) x^*_e +
\sum_{e\in\unsafe\cap\delta(S)} x^*_e \geq (q+2) \qquad\forall\:
        \emptyset\neq{S}\subsetneq{V}$.

\item[(P4)]
Let $A^{(1)} = \{ e \in  \unsafe~|~x^*_e \ge 2/(\alpha+2) \},$
where $\alpha$ denotes the best approximation ratio known for the $\ASC$ problem.
For any nonempty ${S}\subsetneq{V}$,
if $\sum_{e\in\safe\cap\delta(S)} (q+2) x^*_e +
\sum_{e\in\unsafe\cap\delta(S)} x^*_e \leq 2(q+2)$, then
$$\sum_{e\in E\cap\delta(S) - A^{(1)}} u_e(A^{(1)},S) x^*_e \geq D(A^{(1)},S).$$

\item[(P5)]
For each safe edge $f$,
$\sum_{e\in(\safe-f)\cap\delta(S)} (q+1) x^*_e +
\sum_{e\in\unsafe\cap\delta(S)} x^*_e \geq (q+1) \qquad\forall\:
        \emptyset\neq{S}\subsetneq{V}$.

\item[(P6)]
Let $A^{(2)} = \{ e \in  \unsafe~|~x^*_e \ge 1/2 \}.$
For each safe edge $f$ and for any nonempty ${S}\subsetneq{V}$,
if $\sum_{e\in(\safe-f)\cap\delta(S)} (q+1) x^*_e +
\sum_{e\in\unsafe\cap\delta(S)} x^*_e \leq 2(q+1)$, then
$$\sum_{e\in E\cap\delta(S) - A^{(2)}} u_e(A^{(2)},S) x^*_e \geq D(A^{(2)},S).$$
\end{itemize}
\end{lemma}

\begin{proof}
We describe a polynomial-time separation algorithm that identifies
any violation of properties (P3)--(P6). 
Given a vector $x^*$ and a candidate objective value $z$, we first check that $\sum_{e \in E} c_e x_e^* \le z$ (see the discussion on binary search at the end of section~\ref{sec:oneqfgc}). 
If not, we return this as a violated constraint. Otherwise, let $\hG=(\hV,\hE)$
be the capacitated graph where $\hV=V, \hE=E,$ and each edge $e\in\hE$
is assigned a capacity of $u_ex_e^*$.

By following the method of the proof of Lemma~\ref{lem:polytimeKC-oneqfgc},
we can verify that either both (P3)~and~(P4) hold or we can return a violated constraint.

Finally, we consider (P5)~and~(P6).
Consider any safe edge $f$, and the capacitated graph $\hG-f$.
First, we check that the minimum-cut capacity in $\hG-f$ is $\geq(q+1)$
using a polynomial-time minimum-cut algorithm \cite{Schrijver-book}.
If not, we return a minimum cut in $\hG-f$ as a violated constraint.
Otherwise, we know that (P5) is satisfied, and we proceed to verify
(P6) with respect to the set  $A^{(2)}$.

By Karger's results, $\hG-f$ has at most $O(n^4)$ cuts of capacity
$\leq2(q+1)$ (i.e. at most twice the minimum-cut capacity); moreover,
we can enumerate all such cuts in polynomial time \cite{NNI97}.  By
iterating over each of the $O(n^4)$ cuts $\delta(S)$, we can verify
in polynomial~time whether the knapsack-cover inequality for the
non-empty set $S\subsetneq{V}$ and $A^{(2)}\subseteq\unsafe$ is satisfied.
If not, we have found a violated constraint.
\end{proof}
}

The next corollary follows from the above lemma and the well-known fact that 
the ellipsoid algorithm terminates after {$n^{O(1)}$} iterations
of feasibility verification \cite{ellipsoid-book}.
The outer loop of our algorithm runs a binary search for $\lpopt$ 
(see the discussion at the end of section~\ref{sec:oneqfgc}).

\begin{corollary}
There is a polynomial-time algorithm that computes a vector $x^*$ of cost
$c(x^*)\leq\lpopt + \epsilon$ such that $x^*$ satisfies properties~(P3)--(P6),
	for any desired $\epsilon>0$.
Possibly, $x^*$ violates some of the other constraints of ({KCLP:$\,(2,q)$-FGC}).
\end{corollary}

Our algorithm for rounding $x^*$ has two parts.
The first part applies the algorithm of section~\ref{sec:oneqfgc}
and finds a set of edges $\safe_0\cup\unsafe_0$ that is feasible
for the $\oneqplusfgc$ instance.
The second part uses the (assumed) $\rho$-approximation algorithm
for $\twoASC$, as well as Jain's iterative rounding method, and finds
a set of edges $\safe_1\cup\safe_{\alg}\cup{\J}$.
Below, we show that the union of the two sets of edges is feasible
for $\twoqfgc$ and it has cost $\leq(4 (\rho + 1) + \oneqfgcapx)\,c(x^*)$.

Next, we present the details of our algorithm and its analysis.

The first part simply applies the algorithm of section~\ref{sec:oneqfgc}
to round $x^*$ to an edge-set $\safe_0\cup\unsafe_0$ of cost
$\leq\oneqfgcapx\,c(x^*)$ that is feasible for the $\oneqplusfgc$ problem
(described by the inequalities~(1)), where $\safe_0$ is a set of
safe edges and $\unsafe_0$ is a set of unsafe edges.
(We mention that the fact that $x^*$ satisfies the inequalities~(2)
is not relevant for this part.)

In the second part of the algorithm, we start by
defining $\safe_1 = \{e\in\safe: x^*_e\geq 1/4\}$ and
$\unsafe_1 = \{e\in\unsafe: x^*_e\geq 1/2\}$.
Let $\safe_2 = \safe - \safe_1$ and
let $\unsafe_2 = \unsafe - \unsafe_1$.

Define $\C$ to be the family of all sets $S,\emptyset\not=S\subsetneq{V},$ such that
$|\unsafe_1\cap \delta(S)| + \sum_{e\in \unsafe_2 \cap\delta(S)} 2x^*_e < (q+1)$.
For any set $S\in\C$, property~(P3) implies that $\delta(S)$ has at least one safe edge
(otherwise, $\sum_{e\in\unsafe\cap\delta(S)}x^*_e\geq{q+2}$).

{
The analysis of the second part of our algorithm hinges on the next
lemma.  It shows that a (small) constant multiple of the fractional
vector $x^*_{\safe}$ is feasible for the LP~relaxation of the
$\twoASC$ instance defined by $\C$.

\begin{lemma} \label{lem:safepart-twoqfgc}
Let $x^*$ satisfy properties (P3)--(P6) of Lemma~\ref{lem:polytimeKC-twoqfgc}.
A feasible fractional solution to the $\twoASC$ instance defined
by $\C$ is given by the vector $(1_{\safe_1}, 4 x^*_{\safe_2})$; that is,
for any $S\in\C$, we have
$|\safe_1\cap\delta(S)| \:+\: \sum_{e\in \safe_2\cap\delta(S)} 4x^*_e \:\:\geq\:\: 2$.
\end{lemma}

\begin{proof}
We start with a key claim.

\begin{claim}\label{clm:kci-safe1}
Consider any safe edge $f$ and any non-trivial cut $\delta(S),\;S\in\C$.
Then we have
$$\sum_{e\in (\safe - f)\cap \delta(S)} x^*_e \geq 1/2.$$
\end{claim}
\begin{proof}
We prove this claim by examining two cases, based on property~(P6).
First, suppose
$\sum_{e\in(\safe-f)\cap\delta(S)} (q+1) x^*_e +
\sum_{e\in\unsafe\cap\delta(S)} x^*_e > 2(q+1)$.
Then (since $S$ is in $\C$) we have
$$\sum_{e\in\unsafe_1\cap\delta(S)}x^*_e \;+\;
\sum_{e\in\unsafe_2\cap\delta(S)}x^*_e \le
\sum_{e\in \unsafe_1\cap\delta(S)}u_e
\;+\; \sum_{e\in \unsafe_2\cap\delta(S)} {2} u_ex^*_e < (q+1).$$
Hence, $\sum_{e\in(\safe-f)\cap\delta(S)} (q+1) x^*_e \geq (q+1)$,
and it follows that
$\sum_{e\in(\safe-f)\cap\delta(S)} x^*_e \geq 1$.
Thus, the claim holds in the first case.

In the second case, we have
$\sum_{e\in(\safe-f)\cap\delta(S)} (q+1) x^*_e +
\sum_{e\in\unsafe\cap\delta(S)} x^*_e \leq 2(q+1)$.
We will use the fact that $x^*$ satisfies property~(P6).
Let $A = \unsafe_1$.
The knapsack-cover inequality for the non-trivial cut $\delta(S)$ and $A$ is
\[
\sum_{e\in (\safe - f)\cap\delta(S)} u_e(A,S) x_e +
        \sum_{e\in \unsafe_{2}\cap\delta(S)} u_e(A,S) x_e \geq D(A,S),
\]
where $D(A,S) =
        \max\{0,\, q+1-|\unsafe_1\cap\delta(S)|\}$,
$u_e(A,S) = \min\{u_e,\, D(A,S)\}$, each edge $e\in\unsafe_2$ has $u_e=1$,
and each safe edge $e$ has $u_e=q+1$.
Since $S$ is in $\C$, we have the inequality
$|\unsafe_1\cap\delta(S)| + \sum_{e\in \unsafe_{2}\cap\delta(S)} {2} x^*_e < q+1$,
and this inequality can be rewritten in the form
$\sum_{e\in \unsafe_{2}\cap\delta(S)} {2} u_e(A,S) x^*_e < D(A,S)$.
This implies that
$\sum_{e\in \unsafe_{2}\cap\delta(S)} u_e(A,S) x^*_e < \frac{1}{2} D(A,S)$,
and so, by the knapsack-cover inequality,
$\sum_{e\in (\safe-f)\cap\delta(S)} u_e(A,S) x^*_e \geq \frac{1}{2} D(A,S)$.
Since $u_e(A,S) \leq D(A,S)$, we have
$\sum_{e\in (\safe-f)\cap\delta(S)} D(A,S) x^*_e \geq \frac{1}{2} D(A,S)$, hence,
$\sum_{e\in (\safe-f)\cap\delta(S)} x^*_e \geq \frac{1}{2}$.
The claim follows.
\end{proof}

Let us fix any non-trivial cut $\delta(S),\;S\in\C$.
We will prove the lemma by examining two cases.
First, suppose $|\safe_1\cap\delta(S)| \geq 2$. Then the inequality
is trivially true. Second, suppose $|\safe_1\cap\delta(S)| \leq 1$.
Then, Claim~\ref{clm:kci-safe1}
above implies that $\sum_{e\in \safe_2\cap\delta(S)} 4x^*_e\geq2$.
This proves the lemma.
\end{proof}
}

By Lemma~\ref{lem:safepart-twoqfgc}, we can use the $\rho$-approximation
algorithm for $\twoASC$ to find a set of safe edges, call this set
$\safe_{\alg}$, that two-covers all the cuts $\delta(S),\;S\in\C$, incurring a
cost of $\leq4\,\rho\, c(x^*_{\safe})$.

Now, for every non-trivial cut $\delta(S)$,
we either have $|\delta(S)\cap\safe_{\alg}|\geq2$, or we have
$|\delta(S)\cap\unsafe_1| + \sum_{e\in \unsafe_2 \cap\delta(S)} 2x^*_e \geq (q+1)$.

Next, we formulate an $f$-connectivity problem such that
a feasible solution to this problem will cover the non-trivial
cuts $\delta(S)$ with $|\delta(S)\cap\safe_{\alg}|<2$.

For a non-trivial cut $\delta(S)$ define,
\[
f(S) = \begin{cases}
    q+1, &\text{ if } \emptyset\neq{S}\subsetneq{V} \text{ and } |\delta(S)\cap \safe_{\alg}| \leq 1 \\
	0, &\text{otherwise}.
\end{cases}
\]

Observe that $(1_{\unsafe_1}, 2x^*_{\unsafe_2})$ is a fractional
feasible solution to the $f$-connectivity problem.
By Proposition~\ref{prop:wFsupmod}, Jain's iterative rounding
algorithm finds a feasible solution $\J\subseteq{\unsafe}$
to this $f$-connectivity problem of cost
$\leq 2 c({\unsafe_1}) + 2\cdot{2} c(x^*_{\unsafe_2})
 \leq 4 c(x^*_{\unsafe_1}) + 4 c(x^*_{\unsafe_2})
 \leq 4c(x^*_{\unsafe})$.

Finally, we claim that the union of the edge-sets found by the two
parts of the algorithm, namely, $E_{\alg}=(\safe_0\cup\unsafe_0)
\bigcup (\safe_1\cup\safe_{\alg}\cup\unsafe_1\cup{\J})$ is a feasible
solution of the $\twoqfgc$ instance, that is, every non-trivial cut
either has two safe edges (of $E_{\alg}$) or has $q+2$ edges (of $E_{\alg}$).
Consider any non-trivial cut $\delta(S)$ with at most one safe edge (of $E_{\alg}$).
If $\delta(S)\cap{E_{\alg}}$ has one safe edge, then $|\delta(S)\cap{\J}|\geq{q+1}$
(since $\J$ is feasible for the $f$-connectivity problem).
If $\delta(S)\cap{E_{\alg}}$ has no safe edges, then $|\delta(S)\cap{\unsafe_0}|\geq{q+2}$
(since $\safe_0\cup\unsafe_0$ is feasible for the $\oneqplusfgc$ problem).

We claim that the cost of $E_{\alg}$ is
$\leq (4\, (\rho+1) + \oneqfgcapx)\, c(x^*)$, assuming $\rho\geq1$.
To see this,
recall that $x^*_e\geq1/4,\forall{e\in\safe_1}$, and
$x^*_e\geq1/2,\forall{e\in\unsafe_1}$;
moreover, we have
$c(\safe_0\cup\unsafe_0) \leq \oneqfgcapx\, c(x^*)$,~
$c(\safe_1) \leq 4\, c(x^*_{\safe_1})$,~
$c(\safe_{\alg}) \leq 4\, \rho\, c(x^*_{\safe})$,~
$c(\unsafe_1) \leq 2\, c(x^*_{\unsafe_1})$, and
$c(J) \leq 4\, c(x^*_{\unsafe})$.

\twoqFGCreduction*

}

\medskip

\subsection{Reducing \texorpdfstring{$\twoASC$}{2-Cover Small Cuts} to \texorpdfstring{$\twoqfgc$}{(2,q)-FGC} with approximation ratio \texorpdfstring{$O(1)$}{O(1)}}
{
In this subsection, we show that a $\beta$-approximation algorithm
for $(2,q)$-FGC implies a $(\beta+2)$-approximation algorithm for
unit-capacity $\twoASC$. 

Let $\tG=(\tV,\tE)$, $u$, $L$, $c$, $\capbound$ be an instance of
the unit-capacity $\twoASC$ problem; thus, each edge $e\in\tE$ has $u_e=1$.
The goal is to find a cheapest set of links that two-covers all the
non-trivial cuts $\delta(S)$ such that $|\delta(S)\cap\tE| <\capbound$.
Let $L^*$ denote an optimal solution, and let $\opt=c(L^*)$ denote its cost.

We map the above instance of $\twoASC$ to an instance of $\twoqfgc$
(in general, the two instances are not equivalent since the two problems are different).
In the following discussion, we assume that $\capbound >1$;
this ensures that the parameter $q$ is non-negative (see below);
below, we explain that this assumption can be removed.
The graph is $G=(V, E=\safe\cup\unsafe)$ with  $V=\tV$, $\unsafe=\tE$, and $\safe = L$.
Each unsafe edge has cost zero, and each safe edge has the cost of
the corresponding link (in the $\twoASC$ instance).
We fix the parameter $q$ to be $\lceil\capbound\rceil-2$.
Let us use the same notation for a set of links (of the $\twoASC$ instance)
and the corresponding set of safe edges (of the $\twoqfgc$ instance).

A feasible solution to this $\twoqfgc$ instance picks a subset $\J$
of the safe edges such that $\J$ two-covers all non-trivial cuts
$\delta(S)$ with $|\delta(S)\cap \unsafe| \leq q$, and
$\J$ covers all non-trivial cuts $\delta(S)$ with $|\delta(S)\cap\unsafe|=q+1$.

Observe that $L^*\cup\unsafe$ is a feasible solution (of the $\twoqfgc$ instance),
and it has cost $\opt$.
Hence, our $\beta$-approximation algorithm for $\twoqfgc$ finds
a feasible solution of cost $\leq\beta\,\opt$.
Let $\safe_{\alg}$ denote the set of safe edges picked by our
$\beta$-approximation algorithm.

In order to obtain a feasible solution for the $\twoASC$ instance,
we need to augment $\safe_{\alg}$ by a set of safe edges $\J$ such
that $\J$ covers all non-trivial cuts $\delta(S)$ such that
$|\delta(S)\cap\unsafe|=q+1$ and $|\delta(S)\cap\safe_{\alg}|=1$.
We achieve this subgoal via the following $f$-connectivity problem.
For a cut $\delta(S)$ define,
\[
f(S) = \begin{cases}
    \lceil \capbound \rceil = q+2, &\text{ if } \emptyset\neq{S}\subsetneq{V} \text{ and } |\delta(S)\cap \safe_{\alg}| \leq 1 \\ 0, &\text{ otherwise}.
\end{cases}
\]
The graph for this $f$-connectivity problem is the subgraph $G' =
G - \safe_{\alg} = (V,\unsafe\cup (\safe - \safe_{\alg}))$, and the
edge costs are as in $G$ (that is, each unsafe edge has cost zero
and each safe edge of $G'$ has the cost of the corresponding link
of the $\twoASC$ instance).

We claim that $J^* = \unsafe \cup (L^* - \safe_{\alg})$
is a feasible solution to this $f$-connectivity problem.
We verify this claim by considering any non-trivial cut $\delta(S)$ and
examining three cases:
\begin{itemize}
\item[(i)]
If $|\delta(S)\cap\unsafe| \geq q+2$, then clearly
$|\delta(S)\cap{J^*}|\geq f(S)$ (thus the requirement is satisfied);
\item[(ii)]
If $|\delta(S)\cap\unsafe| \leq q$, then
$|\delta(S)\cap\safe_{\alg}|\geq2$, hence, $f(S)=0$ (thus there is no requirement);
\item[(iii)]
If $|\delta(S)\cap\unsafe| = q+1$, then
either $|\delta(S)\cap\safe_{\alg}|\geq2$ and $f(S)=0$
or $|\delta(S)\cap\safe_{\alg}|=1$ and $f(S)=\lceil\capbound\rceil=q+2$
and $|\delta(S)\cap{J^*}|\geq f(S)$
(because $|\delta(S)\cap{L^*}|\geq2>|\delta(S)\cap\safe_{\alg}|$).
\end{itemize}

By Proposition~\ref{prop:wFsupmod}, Jain's iterative rounding
algorithm finds a feasible solution $\J'\subseteq(\safe-\safe_{\alg})\cup\unsafe$
to this $f$-connectivity problem of cost $\leq 2\opt$.

Observe that $\safe_{\alg}\cup{\J'}$ is a feasible solution to the
$\twoASC$ instance because
any non-trivial cut $\delta(S)$ with
$|\delta(S)\cap\unsafe|=|\delta(S)\cap\tE|\leq\lceil\capbound\rceil-2$ is
two-covered by $\safe_{\alg}$, and any other non-trivial cut $\delta(S)$
with $|\delta(S)\cap\unsafe|=|\delta(S)\cap\tE|<\capbound$ is covered
(once) by $\safe_{\alg}$ and is covered (once) by $\J'-\safe_{\alg}$.

Thus we obtain a feasible solution to the $\twoASC$ instance with
cost $\leq(\beta+2)\,\opt$.

\begin{remark}
Recall that we have an instance of $\twoASC$ with unit capacities.
Above, we assumed that $\capbound >1$.
Hence, this assumption misses the case of $0< \capbound\leq 1$.
We can handle the missing case by taking two copies of each edge in $\tE$,
and then fixing $q\;=\; 2\, \lceil\capbound\rceil - 2 \;=\; 0$.
\end{remark}

\twoASCreduction*

\begin{remark}
We could not extend these reductions to $\pqfgc$ problems for $p>2$;
one difficulty is that we could not design $O(1)$ approximation algorithms for the appropriate $f$-connectivity problems.
In particular, we are not able to show similar reductions between $(3,q)$-FGC and $\threeASC$.
\end{remark}
}
}
%%%%%%%%%%
{
\section*{Acknowledgments}
A preliminary version of this paper is published in the proceedings of the
ICALP~2025 conference, \cite{BCKS:icalp2025}.
We thank the anonymous reviewers and the PC of ICALP~2025 for their comments.
We are grateful to David Aleman Espinosa and Sharat Ibrahimpur for reading
a preliminary version, and for their detailed comments.
}

%%%%%%%%%%
\bibliographystyle{plainurl}
\bibliography{cnd-fgc-arxiv}

\begin{thebibliography}{10}

\bibitem{AHM22}
David Adjiashvili, Felix Hommelsheim, and Moritz M{\"u}hlenthaler.
\newblock {Flexible Graph Connectivity}.
\newblock {\em Mathematical Programming}, 192:409--441, 2022.
\newblock \href {https://doi.org/10.1007/s10107-021-01664-9}
  {\path{doi:10.1007/s10107-021-01664-9}}.

\bibitem{ASZ15}
David Adjiashvili, Sebastian Stiller, and Rico Zenklusen.
\newblock {B}ulk-{R}obust {c}ombinatorial {o}ptimization.
\newblock {\em Math. Program.}, 149(1-2):361--390, 2015.
\newblock \href {https://doi.org/10.1007/s10107-014-0760-6}
  {\path{doi:10.1007/s10107-014-0760-6}}.

\bibitem{B2023}
Ishan Bansal.
\newblock {A Global Analysis of the Primal-Dual Method for Pliable Families}.
\newblock {\em CoRR}, abs/2308.15714, 2024.
\newblock URL: \url{https://arxiv.org/abs/2308.15714v2}, \href
  {https://doi.org/10.48550/ARXIV.2308.15714}
  {\path{doi:10.48550/ARXIV.2308.15714}}.

\bibitem{B:ipco2025}
Ishan Bansal.
\newblock A global analysis of the primal-dual method for edge augmentation
  problems.
\newblock In Nicole Megow and Amitabh Basu, editors, {\em Integer Programming
  and Combinatorial Optimization - 26th International Conference, {IPCO} 2025,
  Baltimore, MD, USA, June 11-13, 2025, Proceedings}, Lecture Notes in Computer
  Science, pages 58--71. Springer, 2025.
\newblock \href {https://doi.org/10.1007/978-3-031-93112-3\_5}
  {\path{doi:10.1007/978-3-031-93112-3\_5}}.

\bibitem{BCKS:icalp2025}
Ishan Bansal, Joe Cheriyan, Sanjeev Khanna, and Miles Simmons.
\newblock Improved approximation algorithms for capacitated network design and
  flexible graph connectivity.
\newblock In Keren Censor{-}Hillel, Fabrizio Grandoni, Jo{\"{e}}l Ouaknine, and
  Gabriele Puppis, editors, {\em 52nd International Colloquium on Automata,
  Languages, and Programming, {ICALP} 2025, Aarhus, Denmark, July 8-11, 2025},
  LIPIcs, pages 20:1--20:20, n.p., 2025. Schloss Dagstuhl - Leibniz-Zentrum
  f{\"{u}}r Informatik.
\newblock URL: \url{https://doi.org/10.4230/LIPIcs.ICALP.2025.20}, \href
  {https://doi.org/10.4230/LIPICS.ICALP.2025.20}
  {\path{doi:10.4230/LIPICS.ICALP.2025.20}}.

\bibitem{BCGI24}
Ishan Bansal, Joseph Cheriyan, Logan Grout, and Sharat Ibrahimpur.
\newblock Improved approximation algorithms by generalizing the primal-dual
  method beyond uncrossable functions.
\newblock {\em Algorithmica}, 86(8):2575--2604, 2024.
\newblock \href {https://doi.org/10.1007/s00453-024-01235-2}
  {\path{doi:10.1007/s00453-024-01235-2}}.

\bibitem{BCHI24}
Sylvia~C. Boyd, Joseph Cheriyan, Arash Haddadan, and Sharat Ibrahimpur.
\newblock Approximation algorithms for flexible graph connectivity.
\newblock {\em Math. Program.}, 204(1):493--516, 2024.
\newblock URL: \url{https://doi.org/10.1007/s10107-023-01961-5}, \href
  {https://doi.org/10.1007/S10107-023-01961-5}
  {\path{doi:10.1007/S10107-023-01961-5}}.

\bibitem{CFLP00}
Robert~D. Carr, Lisa~K. Fleischer, Vitus~J. Leung, and Cynthia~A. Phillips.
\newblock Strengthening integrality gaps for capacitated network design and
  covering problems.
\newblock In {\em Proceedings of the Eleventh Annual ACM-SIAM Symposium on
  Discrete Algorithms}, SODA '00, page 106–115, USA, 2000. Society for
  Industrial and Applied Mathematics.
\newblock URL: \url{http://dl.acm.org/citation.cfm?id=338219.338241}.

\bibitem{CCKK15}
Deeparnab Chakrabarty, Chandra Chekuri, Sanjeev Khanna, and Nitish Korula.
\newblock {Approximability of Capacitated Network Design}.
\newblock {\em Algorithmica}, 72(2):493--514, 2015.
\newblock \href {https://doi.org/10.1007/s00453-013-9862-4}
  {\path{doi:10.1007/s00453-013-9862-4}}.

\bibitem{CJ23}
Chandra Chekuri and Rhea Jain.
\newblock {Approximation Algorithms for Network Design in Non-Uniform Fault
  Models}.
\newblock In {\em Proceedings of the 50th International Colloquium on Automata,
  Languages, and Programming}, volume 261, article 36, pages 1--20, 2023.
\newblock \href {https://doi.org/10.4230/LIPIcs.ICALP.2023.36}
  {\path{doi:10.4230/LIPIcs.ICALP.2023.36}}.

\bibitem{DKK22}
Michael Dinitz, Ama Koranteng, and Guy Kortsarz.
\newblock {Relative Survivable Network Design}.
\newblock In Amit Chakrabarti and Chaitanya Swamy, editors, {\em Approximation,
  Randomization, and Combinatorial Optimization. Algorithms and Techniques,
  {APPROX/RANDOM} 2022, September 19-21, 2022, University of Illinois,
  Urbana-Champaign, {USA} (Virtual Conference)}, volume 245 of {\em LIPIcs},
  pages 41:1--41:19. Schloss Dagstuhl - Leibniz-Zentrum f{\"{u}}r Informatik,
  2022.
\newblock \href {https://doi.org/10.4230/LIPIcs.APPROX/RANDOM.2022.41}
  {\path{doi:10.4230/LIPIcs.APPROX/RANDOM.2022.41}}.

\bibitem{DKKN23}
Michael Dinitz, Ama Koranteng, Guy Kortsarz, and Zeev Nutov.
\newblock {Improved} {Approximations} for {Relative} {Survivable} {Network}
  {Design}.
\newblock In Jaroslaw Byrka and Andreas Wiese, editors, {\em Approximation and
  Online Algorithms - 21st International Workshop, {WAOA} 2023, Amsterdam, The
  Netherlands, September 7-8, 2023, Proceedings}, volume 14297 of {\em Lecture
  Notes in Computer Science}, pages 190--204. Springer, 2023.
\newblock \href {https://doi.org/10.1007/978-3-031-49815-2\_14}
  {\path{doi:10.1007/978-3-031-49815-2\_14}}.

\bibitem{GGPSTW94}
Michel~X. Goemans, Andrew~V. Goldberg, Serge~A. Plotkin, David~B. Shmoys,
  {\'{E}}va Tardos, and David~P. Williamson.
\newblock {Improved Approximation Algorithms for Network Design Problems}.
\newblock In {\em Proceedings of the 5th Symposium on Discrete Algorithms},
  pages 223--232, 1994.
\newblock URL: \url{http://dl.acm.org/citation.cfm?id=314464.314497}.

\bibitem{ellipsoid-book}
Martin Grötschel, László Lovász, and Alexander Schrijver.
\newblock {\em Geometric Algorithms and Combinatorial Optimization}, volume~2
  of {\em Algorithms and Combinatorics}.
\newblock Springer Berlin, 1993.
\newblock \href {https://doi.org/10.1007/978-3-642-78240-4}
  {\path{doi:10.1007/978-3-642-78240-4}}.

\bibitem{HLMZ:stacs2025}
Felix Hommelsheim, Zhenwei Liu, Nicole Megow, and Guochuan Zhang.
\newblock Protecting the connectivity of a graph under non-uniform edge
  failures.
\newblock In Olaf Beyersdorff, Michal Pilipczuk, Elaine Pimentel, and Kim~Thang
  Nguyen, editors, {\em 42nd International Symposium on Theoretical Aspects of
  Computer Science, {STACS} 2025, Jena, Germany, March 4-7, 2025}, LIPIcs,
  pages 51:1--51:21. Schloss Dagstuhl - Leibniz-Zentrum f{\"{u}}r Informatik,
  2025.
\newblock URL: \url{https://doi.org/10.4230/LIPIcs.STACS.2025.51}, \href
  {https://doi.org/10.4230/LIPICS.STACS.2025.51}
  {\path{doi:10.4230/LIPICS.STACS.2025.51}}.

\bibitem{HJS:esa2024}
Dylan Hyatt{-}Denesik, Afrouz~Jabal Ameli, and Laura Sanit{\`{a}}.
\newblock {I}mproved {A}pproximations for {F}lexible {N}etwork {D}esign.
\newblock In Timothy~M. Chan, Johannes Fischer, John Iacono, and Grzegorz
  Herman, editors, {\em 32nd Annual European Symposium on Algorithms, {ESA}
  2024, September 2-4, 2024, Royal Holloway, London, United Kingdom}, volume
  308 of {\em LIPIcs}, pages 74:1--74:14. Schloss Dagstuhl - Leibniz-Zentrum
  f{\"{u}}r Informatik, 2024.
\newblock \href {https://doi.org/10.4230/LIPIcs.ESA.2024.74}
  {\path{doi:10.4230/LIPIcs.ESA.2024.74}}.

\bibitem{IV:ipco2025}
Sharat Ibrahimpur and L{\'{a}}szl{\'{o}}~A. V{\'{e}}gh.
\newblock {A}n ${O}(\log {n})$-{A}pproximation {A}lgorithm for
  $(p,q)$-{F}lexible {G}raph {C}onnectivity via {I}ndependent {R}ounding.
\newblock In Nicole Megow and Amitabh Basu, editors, {\em Integer Programming
  and Combinatorial Optimization - 26th International Conference, {IPCO} 2025,
  Baltimore, MD, USA, June 11-13, 2025, Proceedings}, Lecture Notes in Computer
  Science, pages 312--325, n.p., 2025. Springer.
\newblock \href {https://doi.org/10.1007/978-3-031-93112-3\_23}
  {\path{doi:10.1007/978-3-031-93112-3\_23}}.

\bibitem{Jain01}
Kamal Jain.
\newblock {A Factor 2 Approximation Algorithm for the Generalized Steiner
  Network Problem}.
\newblock {\em Combinatorica}, 21(1):39--60, 2001.
\newblock \href {https://doi.org/10.1007/s004930170004}
  {\path{doi:10.1007/s004930170004}}.

\bibitem{Karger93}
David~R. Karger.
\newblock Global min-cuts in {RNC}, and other ramifications of a simple min-cut
  algorithm.
\newblock In Vijaya Ramachandran, editor, {\em Proceedings of the Fourth Annual
  {ACM/SIGACT-SIAM} Symposium on Discrete Algorithms, 25-27 January 1993,
  Austin, Texas, {USA}}, pages 21--30. {ACM/SIAM}, 1993.
\newblock URL: \url{http://dl.acm.org/citation.cfm?id=313559.313605}.

\bibitem{LRS-book}
Lap~Chi Lau, R.~Ravi, and Mohit Singh.
\newblock {\em {Iterative} {Methods} in {Combinatorial} {Optimization}}.
\newblock Cambridge Texts in Applied Mathematics. Cambridge University Press,
  Cambridge, UK, 2011.
\newblock \href {https://doi.org/10.1017/CBO9780511977152}
  {\path{doi:10.1017/CBO9780511977152}}.

\bibitem{NNI97}
Hiroshi Nagamochi, Kazuhiro Nishimura, and Toshihide Ibaraki.
\newblock {Computing All Small Cuts in an Undirected Network}.
\newblock {\em SIAM Journal on Discrete Mathematics}, 10(3):469--481, 1997.
\newblock \href {https://doi.org/10.1137/S0895480194271323}
  {\path{doi:10.1137/S0895480194271323}}.

\bibitem{N:waoa2024}
Zeev Nutov.
\newblock {Improved Approximation Algorithms for Covering Pliable Set Families
  and Flexible Graph Connectivity}.
\newblock In Marcin Bienkowski and Matthias Englert, editors, {\em
  Approximation and Online Algorithms - 22nd International Workshop, {WAOA}
  2024, Egham, UK, September 5-6, 2024, Proceedings}, volume 15269 of {\em
  Lecture Notes in Computer Science}, pages 151--166. Springer, 2024.
\newblock \href {https://doi.org/10.1007/978-3-031-81396-2\_11}
  {\path{doi:10.1007/978-3-031-81396-2\_11}}.

\bibitem{N:mfcs2025}
Zeev Nutov.
\newblock {Tight Analysis of the Primal-Dual Method for Edge-Covering Pliable
  Set Families}.
\newblock In Pawel Gawrychowski, Filip Mazowiecki, and Michal Skrzypczak,
  editors, {\em 50th International Symposium on Mathematical Foundations of
  Computer Science, {MFCS} 2025, Warsaw, Poland, August 25-29, 2025}, LIPIcs,
  pages 82:1--82:14. Schloss Dagstuhl - Leibniz-Zentrum f{\"{u}}r Informatik,
  2025.
\newblock URL: \url{https://doi.org/10.4230/LIPIcs.MFCS.2025.82}, \href
  {https://doi.org/10.4230/LIPICS.MFCS.2025.82}
  {\path{doi:10.4230/LIPICS.MFCS.2025.82}}.

\bibitem{Schrijver-book}
Alexander Schrijver.
\newblock {\em {Combinatorial} {Optimization:} {Polyhedra} and {Efficiency}},
  volume~24 of {\em Algorithms and Combinatorics}.
\newblock Springer, Berlin Heidelberg New York, 2003.

\bibitem{S2025}
Miles Simmons.
\newblock Cover {S}mall {C}uts and {F}lexible {G}raph {C}onnectivity
  {P}roblems.
\newblock Master's thesis, University of Waterloo, Waterloo, ON, Canada,
  September 2025.
\newblock URL: \url{https://hdl.handle.net/10012/22426}.

\bibitem{SBC2025}
Miles Simmons, Ishan Bansal, and Joe Cheriyan.
\newblock {A Bad Example for Jain's Iterative Rounding Theorem for the Cover
  Small Cuts Problem}.
\newblock {\em CoRR}, abs/2504.13105, 2025.
\newblock URL: \url{https://arxiv.org/abs/2504.13105}, \href
  {https://doi.org/10.48550/ARXIV.2504.13105}
  {\path{doi:10.48550/ARXIV.2504.13105}}.

\bibitem{SBC-5approx}
Miles Simmons, Ishan Bansal, and Joe Cheriyan.
\newblock {A 5-Approximation Analysis for the Cover Small Cuts Problem}.
\newblock {\em CoRR}, abs/2602.01462, 2026.
\newblock URL: \url{https://doi.org/10.48550/arXiv.2602.01462}, \href
  {https://arxiv.org/abs/2602.01462} {\path{arXiv:2602.01462}}, \href
  {https://doi.org/10.48550/ARXIV.2602.01462}
  {\path{doi:10.48550/ARXIV.2602.01462}}.

\bibitem{WGMV95}
David~P. Williamson, Michel~X. Goemans, Milena Mihail, and Vijay~V. Vazirani.
\newblock {A Primal-Dual Approximation Algorithm for Generalized Steiner
  Network Problems}.
\newblock {\em Combinatorica}, 15(3):435--454, 1995.
\newblock \href {https://doi.org/10.1007/BF01299747}
  {\path{doi:10.1007/BF01299747}}.

\end{thebibliography}
%%%%%%%%%%

\newpage

{
\appendix

\section{Weakly \texorpdfstring{$\F$}{F}-supermodular functions and Jain's iterative rounding algorithm \label{sec:jain-extension}}

In this section, we review Jain's iterative rounding
algorithm~\cite{Jain01} for network design problems with uniform
edge capacities, along with some extensions that will be used in
section~\ref{sec:twoqfgc}.

{
In the context of approximation algorithms, (several) connectivity
augmentation problems can be formulated in a general framework
called $f$-connectivity. In this problem, we are given an undirected
graph $G = (V,E)$ on $n$ nodes with nonnegative costs $c \in \Qp^E$
on the edges and a requirement function $f:2^V\to\Zint$ on subsets of nodes.
We are interested in finding an edge-set $\J \subseteq E$ with minimum
cost $c(\J) := \sum_{e \in \J} c_e$ such that for all non-trivial cuts
$\delta(S),\ S \subset V$, we have
$|\delta(S) \cap \J| \geq f(S)$.
This problem can be formulated as the following integer program
where the binary variable $x_e$ models the inclusion of the edge $e$ in $\J$:
\begin{align*} \tag{IP: $f$-connectivity}
\label{eq:fconnectivityIP}
\min \quad \qquad		& \quad \sum_{e \in E} c_e x_e 		& \\
\text{subject to: } 	& \quad x( \delta(S) ) \geq f(S) 	& \forall \, \, S \subseteq V \\
					& \quad x_e \in \{0,1\} & \forall \, \, e \in E. \\
\end{align*}
Assuming that the function $f$ is weakly supermodular, {integral,
and has a positive value for some $S\subset{V}$}, Jain \cite{Jain01}
presented a 2-approximation algorithm for the $f$-connectivity problem.
Recently, Dinitz et al.\ \cite{DKK22,DKKN23} discussed extensions of Jain's method to
the setting of locally weakly supermodular functions and weakly
{$\F$}-supermodular functions.  In section~\ref{sec:twoqfgc}, we
use the latter notion.

Let $\F$ be a family of subsets of $V$, such that $\emptyset,V \notin \F$.
A function $f$ is called \textit{weakly $\F$-supermodular} if
{ $f(S)=0,\,\forall{S\not\in\F}$, and } for
all $A,B \in \F$, at least one of the following holds:
\begin{itemize}
    \item $A - B, B - A \in \F$ and $f(A) + f(B) \leq f(A - B) + f(B - A)$
    \item $A \cap B, A \cup B \in \F$ and $f(A) + f(B) \leq f(A \cap B) + f(A \cup B)$
\end{itemize}

\begin{remark}
The reader can follow our presentation independently of \cite{DKKN23}; nevertheless, we mention that \cite[Appendix~D.1]{DKKN23} has a few typos, and there is confusion pertaining to the functions $f$ and $f_{\F}$. Our presentation does not use the notion of $f_{\F}$ and our definition of a weakly $\F$-supermodular function $f$ adds the condition $f(S)=0,\,\forall{S\not\in\F}$.
\end{remark}

The following result is an extension of \cite[Theorem~2.5]{Jain01}.

\begin{proposition}
Let $G$ be a graph and let $x\in\{0,1\}^{E(G)}$ denote a subgraph of $G$.
Let $f: 2^{V(G)} \rightarrow \Zint$ be a weakly $\F$-supermodular function.
Then $f(S) - |\delta_x(S)|$ is also a weakly $\F$-supermodular function.
\end{proposition}

Let us call a node~set $S$ \textit{tight} if $x(\delta(S)) = f(S)$,
and let $\chi^S$ denote the incidence vector of $\delta(S)$.
\cite[Lemma~4.1]{Jain01} states that if $A,B \subseteq V(G)$ are tight sets,
then either $A-B, B-A$ are tight sets and
$\chi^A + \chi^B = \chi^{A-B} + \chi^{B-A}$, or 
$A\cap{B}, A\cup{B}$ are tight sets and
$\chi^A + \chi^B = \chi^{A\cap{B}} + \chi^{A\cup{B}}$.
It can be verified that \cite[Lemma~4.1]{Jain01} holds when the function $f$
is a weakly $\F$-supermodular function.

The other relevant lemmas and theorems of \cite{Jain01} continue
to hold for a weakly $\F$-supermodular function $f$; in particular,
\cite[Theorem~3.1]{Jain01} holds; it states that for any basic
solution $x$ of the LP relaxation of the $f$-connectivity problem,
there is an edge $e$ such that $x_e\geq{1/2}$.

In section~\ref{sec:twoqfgc}, we define requirement functions $f$ of the form
$ \ds
f(S) = \begin{cases}
    k, &\text{ if } \emptyset\neq{S}\subsetneq{V} \text{ and } |\delta(S)\cap E_0| \leq 1 \\ 0, &\text{ otherwise},
\end{cases}
$
where $k$ is a positive integer and $E_0$ is a subset of $E(G)$.
Let $\F$ be the family of nonempty, proper subsets $S$ of $V(G)$
such that $|\delta(S)\cap{E_0}| \leq1$.
For any two sets $A,B\in\F$, either $A-B,B-A\in\F$ or $A\cap{B},A\cup{B}\in\F$;
this can be proved via case analysis;
for example, see the last part of the proof of \cite[Lemma~4.3]{BCHI24}.
Hence, it follows that the above requirement function $f$ is weakly $\F$-supermodular.
This gives the next result.

\begin{proposition} \label{prop:wFsupmod}
Jain's iterative rounding algorithm finds a 2-approximate solution
to an $f$-connectivity problem and a pre-selected edge-set $E_0$
such that $f$ has the form
$\displaystyle f(S) = \begin{cases}
    k, &\text{ if } \emptyset\neq{S}\subsetneq{V} \text{ and } |\delta(S)\cap E_0| \leq 1 \\ 0, &\text{ otherwise}.
\end{cases}
$
\end{proposition}
}
}
%%%%%%%%%%

\end{document}
%)]}